\documentclass[useAMS,usenatbib,usegraphicx]{mn2e}

\usepackage[normalem]{ulem}
\usepackage{color}

\title[Pulsar glitch recovery and the superfluidity coefficients of bulk nuclear matter]{Pulsar glitch recovery and the superfluidity coefficients of bulk nuclear matter}
\author[C. A. van Eysden and A. Melatos]{C. A. van Eysden$^{1}$\thanks{E-mail:~ave@unimelb.edu.au} and A. Melatos$^{1}$\\
$^{1}$School of Physics, University of Melbourne, Parkville, VIC 3010, Australia}
\begin{document}

\date{Accepted XXXX. Received XXXX; in original form XXXX}

\pagerange{\pageref{firstpage}--\pageref{lastpage}} \pubyear{2002}

\maketitle

\label{firstpage}

\begin{abstract}
A two-component hydrodynamic model is constructed of the global superfluid flow induced by two-component Ekman pumping during the recovery stage of a glitch.  The model successfully accounts for the quasi-exponential recovery observed in pulsars like Vela and the ``overshoot'' observed in pulsars like the Crab.  By fitting the model to high-resolution timing data, three important constitutive coefficients in bulk nuclear matter can be extracted: the shear viscosity, the mutual friction parameter, and the charged fluid fraction.  The fitted coefficients for the Crab and Vela are compared with theoretical predictions for several equations of state, including the color-flavor locked and two-flavor color superconductor phases of quark matter.
\end{abstract}


\section{Introduction} \label{gosec1}

Rotational glitches in radio pulsars unfold in two distinct stages: an impulsive spin-up event, followed by a quasi-exponential recovery towards steady spin down.  The time-scales for these two stages are very different, suggesting that they involve different physics. The spin up is unresolved by existing radio timing observations but it is known to be less than $\sim$ 40 {\rm s} in objects like Vela, which have been monitored continuously \citep{mcc87}. The recovery typically lasts for days to weeks \citep{won01}.

The spin-up stage is generally attributed to a transfer of angular momentum from the neutron superfluid core to the crustal lattice via superfluid vortex unpinning \citep{and75a}.  Recent data, which reveal that glitch sizes and waiting times follow power-law and Poissonian distributions respectively in individual pulsars \citep{mel08}, suggest that the glitch trigger is an unpinning avalanche in the superfluid vortex array, occurring via a self-organized critical process \citep{war08} or a coherent noise process \citep{mel09}. To understand the collective nature of the trigger, one must synthesize a wealth of condensed matter physics at the microscopic level, including vortex instabilities, the distribution of pinning potentials, and the coupling of the superfluid to the charged fluid by entrainment.
Separately, the spin-up event has been modeled hydrodynamically by averaging over the discrete vortex microphysics \citep{gla09,sid09}.  In the latter class of models, angular momentum is transferred from the superfluid to the charged fluid and hence the crust by mutual friction.

The recovery stage is thought to reflect the restoration of superfluid-lattice corotation by viscous and/or magnetic forces \citep{bay69b,boy72,loh75}.
However, viscosity estimates based on electron-electron scattering and estimates of the magnetic tension predict shorter recovery time-scales than observed \citep{eas79}.
Moreover, although mutual friction estimates based on electron scattering off neutron magnetic moments in vortex cores are consistent with long recovery times \citep{bay69b} scattering off vortex cores magnetized by neutron-proton entrainment \citep{and75a} couples the proton-electron plasma to the neutron superfluid on a time-scale much shorter than observed \citep{alp84}.  In the latter scenario, most of the fluid interior locks to the crust, and the recovery stage is attributed to vortex repinning and creep \citep{alp93,alp96,sid09}.  All the above factors feed into the global hydrodynamics in ways that remain unclear.

Three aspects of the glitch recovery process are truly puzzling from a physical perspective.
First, if the recovery occurs hydrodynamically via the Ekman process \citep{eas79,abn96}, one would naively expect the recovery time-scale to be the same for all glitches in a given pulsar, because the glitch amplitude and post-glitch perturbation of the flow are small and hence the process is linear.
The time-scale for {\rm linear} Ekman pumping is independent of glitch amplitude and depends instead on constitutive properties such as viscosity and mutual friction, which do not vary significantly during the interval between glitches \citep{rei93,abn96}.  Contrary to expectations, however, the recovery time-scales observed in the Crab and Vela pulsars cover wide ranges \citep{mcc87,alp93,alp96,won01,dod02}, and the same seems to be true in less heavily studied pulsars \citep{per07}.

Second, glitch recovery cannot be parameterized by a single exponential decay.
In almost all cases, at least two exponentials are required \citep{mcc87}, and up to four have been fitted to high resolution data \citep{dod02}.
The corresponding time-scales typically range from $0.3\,{\rm d}$ to $300\,{\rm d}$, suggesting that there are multiple physical processes involved.
This behavior is inconsistent with standard Ekman pumping: for example, the spin up of a stratified viscous fluid in a cylinder involves one time-scale \citep{abn96}, and the spin up of an unstratified superfluid between parallel plates involves two \citep{rei93}.  The multiple time-scales are often attributed to the variation of vortex pinning strength (and hence vortex creep rate) with depth and have been modeled by dividing the star's moment of inertia into multiple components \citep{alp93,alp96,sed02}.

Third, although many pulsars recover monotonically and quasi-exponentially like Vela, some pulsars do not.
The Crab pulsar consistently ``overshoots'' during its recovery, decelerating below its steady-state angular velocity before rising again asymptotically \citep{won01}.
This suggests that the vortex unpinning event fails to redistribute angular momentum evenly throughout the star; differential rotation must persist between one or more internal components, which do not achieve corotation simultaneously during the recovery stage.

In this paper, we consider an idealized hydrodynamic model for the recovery stage of pulsar glitches.  Our model consists of a rigid outer crust containing a two-component superfluid.
The viscous component (proton-electron plasma) spins up via Ekman pumping, whereby viscous stresses transfer angular momentum to the fluid in a boundary layer, which is then convected throughout the star by the Coriolis force \citep{gre63,rei93,eas79,abn96}.
The inviscid component (neutron condensate) interacts with the viscous fluid component via the mutual friction force,
which arises from electron scattering off magnetized vortex cores \citep{men91b}.
The important effects of vortex tension and pinning and macroscopic entrainment are neglected to keep the problem tractable analytically.
In contrast to previous studies, the crust has a finite mass and responds to the viscous torque applied by the interior fluid.
This back-reaction in turn modifies the flow field and we calculate this self-consistently.
The magnetic dipole torque is neglected, because the recovery stage is much shorter than the electromagnetic spin-down time.
Our approach differs from previous studies of the recovery stage, which incorporate vortex pinning and creep but do not solve self-consistently for the {\it global} two-fluid flow pattern produced by Ekman pumping \citep{sed02,alp84, alp93,alp96}.
The global flow pattern and viscous back-reaction are built into a hydrodynamic model self-consistently for the first time, yielding a tool that can be used to investigate a variety of pictures for pulsar interiors, as we discuss below.
We do not assume a priori knowledge of any constitutive coefficients, leaving them completely free to be determined by observations.
Thus, we seek to determine what aspects of the post-glitch relaxation a self-consistent hydrodynamic model can and cannot explain, in order to clarify what elements of non-hydrodynamic physics are absolutely required by the data.

In \S\ref{gosec2}, we present an analytic model to describe the rotational evolution of the crust during the recovery stage.
The model is based on a recent analytic solution by \citet{van10} for the spin-up problem of a two-component superfluid in spherical Couette geometry.
In this paper, we specialize to the case where there is no inner core, saving the more general problem for a future study.
The output of the model can be compared directly against high-resolution radio timing data.
It is fitted to the quasi-exponential recovery of Vela glitches in \S\ref{gosec3} and to the ``overshoot'' recovery in Crab glitches in \S\ref{gosec4}.
In both cases, the relevant superfluidity coefficients are extracted.
Finally, in \S\ref{gosec5}, we compare the fitted values of the superfluidity coefficients with theoretical predictions for a standard $^1S_0$ neutron superfluid \citep{cut87, men91b} and for various dissipation channels in an exotic strange-quark superfluid, e.g. in a color flavor locked phase \citep{mad00, man08,alf09}.  There is currently a flowering of interest in dissipative transport processes in bulk nuclear matter, as the foregoing references indicate.  Quantitative glitch recovery studies, especially when performed on pulsars with different ages,  offer a promising way to measure nuclear transport coefficients experientially in the many-body, MeV regime, which is inaccessible at present in terrestrial laboratories.







\section{Spin up of a spherical star: two-fluid theory} \label{gosec2}
In the absence of a satisfactory microscopic explanation of the glitch trigger, we consider the hydrodynamic response of the stellar interior to the following set of idealized initial conditions.
Consider a thin, spherical shell of radius $R$, representing the solid crust of the star, which rotates at angular velocity $\Omega$ just before the glitch $(t=0^-)$.
Suppose the shell contains a two-component superfluid, whose inviscid (Bose-Einstein-condensed neutrons) and viscous (uncondensed neutrons, protons and other charged species) components rotate rigidly but differentially with angular velocities $\Omega_{s0}$ and $\Omega_{n0}\neq\Omega_{s0}$ respectively at $t=0^-$.
Immediately after the glitch, the crust accelerates instantaneously to reach an angular velocity $\Omega+\Delta\Omega$ at $t=0^+$.
Subsequently, at $t>0$, the angular velocity of the crust evolves according to $\Omega+\Delta\Omega f(t)$ [with $f(0^+)$=1], while the two superfluid components participate in coupled Ekman pumping, with the coupling provided by mutual friction.
During the Ekman process, neither component rotates rigidly, unlike in other ``body-averaged'' treatments \citep{sid09}.
At its conclusion, however, the components tend toward corotation.

The above initial conditions, although idealized, are consistent with the spirit of the superfluid vortex unpinning paradigm, in the sense that the inviscid component leads the other components before the glitch and transfers part of its excess angular momentum almost instantaneously (e.g. via Kelvin waves) to the crust during the vortex unpinning avalanche \citep{alp84,mel08}.
The absence of a ``reservoir effect'' (glitch size $\propto$ waiting time since previous glitch) in the data guarantees that the avalanche event nullifies only a small fraction of the accumulated differential rotation and leaves $\Omega_{s0}\neq\Omega_{n0}$ at $t=0^+$ \citep{mel09}.
(The instantaneous decrease in $\Omega_s$ at $t=0$ accompanying the avalanche can be absorbed into $\Omega_{s0}$ without loss of generality.)
By the same token, the final conditions (corotation between all components) are less realistic; they fail to acknowledge that differential rotation must persist at all times to avoid the reservoir effect.
This flaw is shared by all published hydrodynamic models, which elect not to describe the stochastic vortex repinning process.


In what follows, we adopt cylindrical coordinates $(r,\phi,z)$ and work in the non-inertial frame rotating with the pre-glitch angular velocity $\Omega \bmath{k}$ of the crust, where $\bmath{k}$ denotes the unit vector along the $z$-axis.

\subsection{Interior flow} \label{gosec2a}

The flow in the interior of the pulsar is governed by the two-fluid Hall-Vinen-Bekharevich-Khalatnikov (HVBK) equations.  These equations can be linearised on the basis that the glitch amplitude is small $(\Delta\Omega\ll\Omega)$.  In the rotating frame, the incompressible, linearized HVBK equations take the dimensionless form \citep{per08}
\begin{equation} \label{goeq1}
E^{1/2}\frac{\partial \bmath{v}_n}{\partial \tau}+ 2 \bmath{\bmath{k}} \times \bmath{v}_n=-\nabla p_n +E \nabla^2 \bmath{v}_n+\rho_s \bmath{F}\, ,
\end{equation}
\begin{equation}\label{goeq2}
E^{1/2}\frac{\partial \bmath{v}_s}{\partial \tau}+ 2 \bmath{k} \times \bmath{v}_s=-\nabla p_s-\rho_n \bmath{F}\, ,
\end{equation}
\begin{equation} \label{goeq3}
\nabla\cdot\bmath{v}_n=0\, ,
\end{equation}
\begin{equation} \label{goeq4}
\nabla\cdot\bmath{v}_s=0\, .
\end{equation}
The components are coupled by the mutual friction force \citep{hal56b}
\begin{equation} \label{goeq5}
 \bmath{F}=B  \bmath{k}\times\left[\bmath{k}\times\left(\bmath{v}_n-\bmath{v}_s\right)\right] +B' \bmath{k}\times\left(\bmath{v}_n-\bmath{v}_s\right) \,,
\end{equation}
where we assume the flow is laminar [cf. \citet{mel07}] and $\bmath{F}$ takes the anisotropic Hall-Vinen form rather than the isotropic Gorter-Mellink form [i.e., there is no vortex tangle; cf. \citet{per05,per06,and07}].  The symbols $\bmath{v}_{n,s}$ and $p_{n,s}$ denote the velocity and pressure of the inviscid $(s)$ and viscous $(n)$ components.  $B$ and $B'$ are dimensionless parameters which we hope to measure by applying the spin-up model to pulsar timing data.
The velocity and pressure scales are chosen to be $R \Delta \Omega$ and $\rho  \Omega R^2 \Delta \Omega$ respectively.  The mass densities of the viscous and inviscid components, $\rho_n$ and $\rho_s$, are scaled to the total density, $\rho$, so that we have
\begin{equation} \label{goeq6}
 \rho_n+\rho_s=1 \, .
\end{equation}
To scale the time coordinate, we define the Ekman time \citep{gre63}
\begin{equation} \label{goeq7}
\tau= E^{1/2}\Omega t\,
\end{equation}
in terms of the Ekman number
\begin{equation} \label{goeq8}
 E=\frac{\mu}{\rho_n R^2 \Omega}\, ,
\end{equation}
where $\mu$ is the shear viscosity. In a neutron star, $E$ is very small, with $E\la 10^{-10}$ typically \citep{mel07}.
Note that the centripetal force terms are absorbed into the definitions of the pressures, as the flow is incompressible.

Equations (\ref{goeq1})--(\ref{goeq4}) neglect entrainment, whereby the flow of one fluid imparts momentum to the other and vice versa via a quantum mechanical current-current interaction \citep{and06,and08,sid09}.  Entrainment is expected to play an important role in the dynamics of neutron star interiors, but we neglect it in this paper to keep an already difficult problem analytically tractable. [Superfluid spherical Couette flow has never been solved analytically before and was treated numerically only recently; see \citet{per08}.] Similarly, magnetic fields are also neglected, even though \citet{van08} showed that they are important in vortex dynamics, e.g. suppressing the Donnelly--Glaberson instability. Vortex tension is neglected on the basis that the inter-vortex spacing is small, although the exact nature and strength of vortex pinning to the crust is still being debated \citep{lin09}.

The presence of the mutual friction force in (\ref{goeq1})--(\ref{goeq4}) introduces new physics into the traditional Ekman pumping process.
In addition to the classical Ekman time-scale $ E^{-1/2} \Omega^{-1}$, the mutual friction introduces a second time-scale, $B^{-1}\Omega^{-1}$, characterized by the coupling strength.
The ordering of these time-scales governs the dynamics of the spin-up flow.  For $B\sim1$, the inviscid and viscous fluids are locked together; differential rotation is removed by mutual friction over a few rotation periods.
For $B\sim E^{1/2}$, the spin-up time is a combination of $ E^{-1/2} \Omega^{-1}$ and $B^{-1}\Omega^{-1}$; the viscous fluid spins up via Ekman pumping, while ``dragging'' the inviscid component along via the mutual friction.
For  $B\ll E^{1/2}$, Ekman pumping proceeds for the viscous fluid uninhibited by mutual friction over the time-scale $E^{-1/2} \Omega^{-1} $, while the inviscid component is brought into corotation over the much longer time-scale $B^{-1}\Omega^{-1}$.

In helium II, the mutual friction arises when excited states scatter off the vortex lines, and $B$ and $B'$ are of order unity.
In neutron stars, several mechanisms give rise to mutual friction, including: (1) electron scattering off the neutron magnetic moments in vortex cores, with characteristic time-scale $\sim1 \rm{yr}$ \citep{bay69b};  (2) electron scattering off entrained protons,  which magnetize the vortex cores, with time-scale $\sim 1\rm{s}$ \citep{alp84, men91b}; and (3) excited states scattering off vortices like in Helium II.
Processes (1) and (2) bring the neutron superfluid into corotation with the plasma via electromagnetic interactions \citep{alp84}.
Uncharged particle species (e.g. excited states of the neutron condensate or exotic particle species) couple via (3).
Terrestrial experiments on liquid helium reveal that $\sim 10\%$ of the fluid is in excited states even at absolute zero, well above the ideal Bose gas fraction, because of the non-ideal nature of the molecular forces.
Similar non-ideal behavior for the strong nuclear force is not ruled out at present.

Transport coefficients such as viscosity and mutual friction depend on depth (via the density and temperature, see \S\ref{gosec5}), and neutron stars are strongly stratified.  Hence, by treating $E$, $B$ and $B'$ as uniform in this paper (to keep the problem tractable), we are obliged to interpret the values generated by fitting the model to data (see \S\ref{gosec3} and \S\ref{gosec4}) as body-averaged effective values.
The average is taken over regions of the star with different compositions (e.g. exotic quark matter), densities, and temperatures.
Importantly, the average depends sensitively on the depth to which Ekman pumping extends; it is reduced dramatically by compressibility and stratification \citep{abn96,van10}, crucial physics which we do not consider here.

The general analytic solution to (\ref{goeq1})--(\ref{goeq4}) in spherical Couette geometry was derived recently by \citet{van10}, generalizing the boundary layer analysis of \citet{gre63}.
In this paper we restrict our attention to the simple situation where there is no inner core and specialize to the regime $B,B'\ll1$ (which we verify a posteriori in several examples in \S\ref{gosec3} and \S\ref{gosec4}) pertaining to glitching pulsars \citep{men91b,sid09}.
For the general solution with no inner core, the reader is presented the Appendix.
As discussed by \citet{van10}, the solution to (\ref{goeq1})--(\ref{goeq4}) is expressed most neatly in terms of the total mass flux, defined as
\begin{equation}  \label{goeq9}
\bmath{v} =\rho_n \bmath{v}_{n} + \rho_s \bmath{v}_{s}\, .
\end{equation}
The azimuthal component of the total mass flux is given by
\begin{equation} \label{goeq10}
v_{\phi}(r,\tau)=\frac{r \omega_+(r) \omega_-(r)}{\beta\left[\omega_+(r)-\omega_-(r)\right]} \int_0^\tau\rmn{d}\tau'\, f(\tau')
\end{equation}
\[
\phantom{v_{\phi}(r,\tau)=} \times\left\{  \left[\omega_+(r)+\beta\right] e^{\omega_+(r)\left(\tau-\tau'\right)} \right.
\]\[
\phantom{v_{\phi}(r,\tau)=} \left. -\left[\omega_-(r)+\beta \right]e^{\omega_-(r)\left(\tau-\tau'\right)}\right\} \nonumber
\]\[
\phantom{v_{\phi}(r,\tau)=} -\frac{r \Omega_{n0} \omega_+(r) \omega_-(r) }{\beta\left[\omega_+(r)-\omega_-(r)\right]} \left[e^{\omega_+(r) \tau}- e^{\omega_-(r) \tau}\right] \nonumber
\]\[
\phantom{v_{\phi}(r,\tau)=} -\frac{r \Omega_0}{\omega_+(r)-\omega_-(r)}\left[\omega_-(r)e^{\omega_+(r) \tau}-\omega_+(r) e^{\omega_-(r) \tau}\right] \, .
\]
In (\ref{goeq10}), $\Omega_0$ and $\Omega_{n0}$ denote the initial angular velocities of the total mass flux and viscous component in the rotating frame, scaled to $\Delta \Omega$.
The radius-dependent time-scales $\omega_{\pm}^{-1}$ are a mixture of the classical Ekman time-scale and the superfluid mutual friction coupling time-scale as discussed above, with
\begin{equation} \label{goeq11}
  \omega_{\pm}(r)=-\frac{1}{2}\left[\beta+\left(1-r^2\right)^{-3/4}\right]
\end{equation}
\[
\phantom{\omega_{\pm}(r)} \pm \left\{\frac{1}{4}\left[\beta+\left(1-r^2\right)^{-3/4}\right]^2-\beta \rho_n \left(1-r^2\right)^{-3/4}\right\}^{1/2} \, ,
\]
and
\begin{equation} \label{goeq12}
 \beta=B E^{-1/2} \,.
\end{equation}
Note that $B'$ no longer appears in the equations in the regime $B,B'\ll1$.
The boundary conditions leading to (\ref{goeq10}) are that the viscous component $\bmath{v}_n$ satisfies no penetration and no slip at $z=\pm(1-r^2)^{1/2}$, while the inviscid component $\bmath{v}_s$ satisfies no penetration.

We quote only the result for the interior azimuthal flow in (\ref{goeq10}). The full solution also consists of the radial and vertical flows describing the Ekman flow, as well as boundary layer corrections \citep{van10}.  However, the latter elements do not appear explicitly in the theory of glitch recovery.

\subsection{Back reaction torque on crust} \label{gosec2b}
Equation (\ref{goeq10}) gives the solution of (\ref{goeq1})--(\ref{goeq4}) in terms of a general boundary condition $f(\tau)$.  To solve the glitch recovery problem self-consistently, we must determine the time evolution of $f(\tau)$ due to the hydrodynamic torque on the rigid crust.  This is done by integrating the viscous stress tensor over the stellar surface $z=\pm(1-r^2)^{1/2}$. The result is \citep{van10}
\begin{equation} \label{goeq13}
 \frac{d f(\tau)}{d \tau}= -\frac{15 K}{2}\int_{0}^{1}\rmn{d}r\,r^2 \left(1-r^2\right)^{1/2}\frac{\partial v_{\phi}(r,\tau)}{\partial \tau}  \, ,
\end{equation}
where $K$ is the ratio of the moment of inertia of the total fluid to the moment of inertia of the crust.  Equations (\ref{goeq10}) and (\ref{goeq13}) constitute a closed pair of integro-differential equations for the unknowns $v_{\phi}(r,\tau)$ and $f(\tau)$.
Substituting (\ref{goeq10}) into (\ref{goeq13}), we derive an integral equation for the spin evolution of the crust, viz.
\begin{equation}\label{goeq14}
 f(\tau)=-\rho_n K \int_0^\tau \rmn{d}\tau' \, \left[\dot{g}^A(\tau-\tau')+\dot{g}^B(\tau-\tau')\right]f(\tau')
\end{equation}
\[\phantom{ f(\tau)=} + \rho_n K  \left[  g^A(\tau)\Omega_{n0}+ g^B(\tau) \Omega_{0} \right] +1 \, ,
\]
where we define
\begin{equation} \label{goeq15}
 g^A(\tau)=\frac{15}{2}\int_{0}^{1} \rmn{d}r \, \frac{r^3 \left[  e^{\omega_+(r)\tau}-e^{\omega_-(r)\tau} \right]}{\left(1-r^2\right)^{1/4}\left[\omega_+(r)-\omega_-(r)\right]}
\end{equation}
\begin{equation} \label{goeq16}
 g^B(\tau)=\beta \int_0^\tau \rmn{d}\tau' \, g^A(\tau') \, .
\end{equation}
It is straightforward to solve (\ref{goeq14}) for $f(\tau)$ numerically, by guessing an initial trial function for $f(\tau)$ (e.g., exponential), substituting it into the right-hand side of (\ref{goeq14}), updating $f(\tau)$ via  underrelaxation \citep{press}, and iterating.


The parameter $K$ can be interpreted in a number of ways, depending on the glitch microphysics.
It can represent the moment of inertia of the crust (i.e., ionic lattice alone), or the total moment of inertia of the crust and all components of the star coupled to it on a very short time-scale.
Broadly speaking, if the crust is strongly coupled magnetically to the proton-electron plasma \citep{bay69b,eas79} and subsequently to other components of the star, via electron scattering off magnetized neutron vortices \citep{alp84} or vortex-fluxoid interactions \citep{van08}, one has $K\sim1$.  The viscous fluid component then represents excited neutron states (non-ideal internuclear forces) or other neutral exotic particle species decoupled magnetically from the crust (see \S\ref{gosec4e}).
Alternatively, if $K$ refers just to the ionic lattice, one has $K\ga50$ (see \S\ref{gosec5}).
\footnote{In a more general theory where the magnetic dipole spin-down torque is included, it would act on the crust and any components strongly coupled to it.}

An advantage of the above model is that is predicts the behavior of an observable, namely $\Omega+\Delta\Omega f(\tau)$, the post-glitch angular velocity of the crust.  The model has {\it six} free parameters: $\rho_n$, $K$, $B$, $E$, $\Omega_{n0}$ and $\Omega_0$.

\subsection{Fitting to radio timing data} \label{gosec2c}
It is customary to fit the post-glitch crustal frequency $\nu_g(t)$  {\it empirically} using a sum of $N$ exponentials, with amplitudes and $e$-folding times $\Delta\nu_n$ and $t_n$ respectively, i.e.,
\begin{equation} \label{goeq17}
 \nu_g(t)=\nu_{g0} +\Delta\nu_p+\sum_{n=1}^N \Delta\nu_n e^{-t/t_n} \, ,
\end{equation}
where $\Delta\nu_p$ is the permanent part of the spin up, $\nu_{g0}=\nu_g(0^-)$ is the spin frequency just before the glitch, and one has $N \le 4$ typically \citep{mcc87,alp93,alp96,won01}.  Normalizing (\ref{goeq17}) to the observed frequency jump
\begin{equation} \label{goeq18}
 \Delta \nu=\Delta\Omega/2\pi=\Delta\nu_p+\sum_{n=1}^N \Delta\nu_n
\end{equation}
immediately after the glitch, and writing $t$ in terms of the Ekman time (\ref{goeq8}), we obtain
\begin{equation} \label{goeq19}
 f_{obs}(\tau)=\frac{\Delta\nu_p}{\Delta\nu}+\sum_{n=1}^N \frac{\Delta\nu_n }{\Delta\nu} \exp \{-\tau/[E^{1/2} 2\pi\nu_{g0} t_n]\}\, .
\end{equation}
Our goal is to adjust the six model parameters $\rho_n$, $K$, $B$, $E$, $\Omega_{n0}$ and $\Omega_0$ until the analytic solution $f(\tau)$ from (\ref{goeq14}) matches the observed behavior $f_{obs}(\tau)$ as accurately as possible.

Post-glitch timing typically yields $\sim 10^2$ data points ($\sim 1$ per day) during the recovery stage, with formal residuals $\la 1\%$.  Hence the six model parameters are overconstrained in principle.  In practice, the comparison of $f_{obs}(\tau)$ and $f(\tau)$ is done by eye, and the idealized theory in \S\ref{gosec2a} and \S\ref{gosec2b} is best matched to the approximate observed behavior given by (\ref{goeq17}).

Two conditions on the model parameters can be read off the data straight away.  First, conservation of angular momentum in the steady state implies \citep{van10}
\begin{equation} \label{goeq20}
\frac{\Delta\nu_p}{\Delta\nu}=f(\infty)=\frac{1+K \Omega_0}{1+K} \, .
\end{equation}
Second, the initial slope $\dot{f}(0^+)$ satisfies \citep{van10}
\begin{eqnarray} \label{goeq21}
 -\frac{\sum_{n=1}^N \Delta \nu_n/t_n}{2\pi \nu_{g0} \Delta\nu} =E^{1/2} \dot{f}(0^+)=-\frac{20}{7}\rho_n K E^{1/2} \left(1-\Omega_{n0}\right)\, .
\end{eqnarray}
Equations (\ref{goeq20}) and (\ref{goeq21}) provide two conditions on the model variables (on the right-hand sides) in terms of the measured quantities $\Delta\nu_p,\Delta\nu_n,t_n,\nu_{g0}$ (on the left-hand sides).  In general, the remaining four conditions must be determined through trial and error by matching $f_{obs}(\tau)$ and $f(\tau)$ by eye.

\subsection{An approximate solution involving dual exponentials} \label{gosec2d}
If the superfluidity coefficients satisfy $E^{1/2}\ll B,B'\ll1$ (which we verify a posteriori in several examples in \S\ref{gosec3} and \S\ref{gosec4}) and $K\gg1$, then (\ref{goeq14}) has the approximate solution
\begin{equation}  \label{goeq22}
 f(\tau)=\left[1-C-f(\infty)\right]\exp[{-(20/7)\rho_n(1+K)\tau}]
\end{equation}
\[\phantom{f(\tau)=} +C \exp({-B E^{-1/2} \tau})+f(\infty)\, ,
\]
where $f(\infty)$ is defined in (\ref{goeq20}) and $C$ is given by
\begin{equation} \label{goeq23}
 C=\frac{20 \rho_n K \left(\Omega_0-\Omega_{n0}\right)}{7 B E^{-1/2}-20\rho_n K } \, .
\end{equation}
This approximation matches the exact numerical solution to $\leq1\%$ for $K \geq 10^3$.
Comparing (\ref{goeq19}) and (\ref{goeq22}), we can read off the model parameters directly in terms of $t_1$ and $t_2$:
\begin{equation} \label{goeq24}
 B=\left(2\pi\nu_{g0} t_1\right)^{-1} \, ,
\end{equation}
\begin{equation} \label{goeq25}
 \frac{20}{7}E^{1/2}\rho_n(1+K)=\left(2\pi\nu_{g0} t_2\right)^{-1} \, ,
\end{equation}
Equations (\ref{goeq24}) and (\ref{goeq25}) provide two more observational constraints in addition to (\ref{goeq20}) to (\ref{goeq21}) for the model parameters.
The model is therefore underconstrained; we are free to choose two parameters.  For any given $K$ and $\rho_n$, say, equations (\ref{goeq20}), (\ref{goeq21}), (\ref{goeq24}) and  (\ref{goeq25}) determine $B$, $E$, $\Omega_{n0}$ and $\Omega_0$.

Note that a second, equally valid set of model parameters can be extracted by taking the time-scales the other way around, i.e. by swapping $t_1$ and $t_2$ in (\ref{goeq24}) and (\ref{goeq25}).  The approximate solution also provides an excellent initial trial function when iterating (\ref{goeq14}).
Of course, the full timing ephemeris contains more information than equation (\ref{goeq17}) with $N=2$, so the model is actually overconstrained in principle, as noted above.  Nevertheless, the quality of the data and realism of the model are such that two of the parameters are free in practice in many objects.  If we insist that $K$ and $\rho_n$ (say) do not change from one glitch to the next in an individual pulsar, then we can constrain all six model parameters uniquely.  This is done in \S\ref{gosec3} and \S\ref{gosec4}.

\section{Vela}  \label{gosec3}

\subsection{Data} \label{gosec3a}

\begin{table*}
 \centering
 \begin{minipage}{1500mm}
  \caption{Timing parameters for large glitches $(\Delta\nu\geq1\rm{\mu Hz})$ in the Vela pulsar $\left(\nu_{g0}=11.2\, {\rm Hz}\right)$.  An asterisk in the CM column denotes continuous monitoring.}
  \begin{tabular}{@{}clrrrrrrrrrrrrc@{}}
  \hline
   Glitch & Date & MJD & CM & $t_4$ & $t_3$ & $t_2$ & $t_1$ & $\Delta\nu_4$ & $\Delta\nu_3$ & $\Delta\nu_2$ & $\Delta\nu_1$ & $\Delta\nu_p$ & $\Delta\nu$ & Ref. \\
    & & &  & \multicolumn{4}{c}{{\rm d}} & \multicolumn{6}{c}{$\mu${\rm Hz}} &  \\
 \hline
1	& 28-Feb-1969	& 40280	& &	&	& 10	& 120 &	&	& 0.052		& 0.467	& 25.6	& 26.2 & 1 \\
2	& 29-Aug-1971	& 41192	& &	&	& 4	& 94 &	&	& 0.036		& 0.30	& 22.6	& 22.9 & 1 \\
3	& 09-Sep-1975	& 42664	& &	&	& 4	& 35 &	&	& 0.01		& 0.079	& 22.2	& 22.2 & 1 \\
4	& 13-Jul-1978	& 43693	& &	&	& 6	& 75 &	&	& 0.083		& 0.389	& 33.8	& 34.3 & 1\\
5	& 10-Oct-1981	& 44888	& &	&	& 6	& 14 &	&	& 0.01		& 0.024	& 12.7	& 12.7 & 1\\
	& 10-Oct-1981	& 44889	& & 	& 	& 1.6	& 233 &	&	& 0.092		& 2.26	& 10.5	& 12.8 & 2\\
6	& 10-Aug-1982	& 45192	& &	&	& 3	& 21.5 & &	& 0.057		& 0.126	& 22.8	& 23.0 & 1\\
	& 10-Aug-1982	& 45192	& & 	& 	& 3.2	& 60 &	&	& 0.23		& 0.79	& 22.0	& 23.1 & 2\\
7	& 12-Jul-1985	& 46258	& & 	&	& 6.5	& 332 &	&	& 0.066		& 2.76	& 15.1	& 17.9 & 2\\
	& 12-Jul-1985	&	& &	&	& 6.8	& &	&	& 0.165		& 	& 		& 	  & 3\\
8	& 24-Dec-1988	& 47519	&*&	& 0.4	& 4	& 96 &	& 0.108	 & 0.086	& 0.376	& 19.7	& 20.2 & 4\\
	& 24-Dec-1988	& 47520	&*& 	& 0.73	& 6.97	& 707 &	& 0.092	 & 0.083	& 6.74	& 13.3	& 20.2 & 5\\
9	& 20-Jul-1991	& 48458	&*&	& 0.56	& 5.94	& 254 &	& 0.255 & 0.169		& 2.84	& 27.1	& 30.3 & 5\\
	& 20-Jul-1991	& 	&*& 	& 0.59	& 4.9	& 49 &	& 0.317 & 0.152		& 0.231 &		&	  & 3\\
10	& 26-Jul-1994	& 49560	&*&	&	&	& &	&	&		& 	& 9.6	& 9.6  & 5\\
	& 26-Jul-1994	&	&*&	&	&	& &	&	&		&	&		& 	  & 3\\
11	& 27-Aug-1994	& 49592	&*&	&	& 1.59	& 15 &	&	& 0.024	& 0.027	& 2.1	& 2.2 & 5\\
	& 27-Aug-1994	& 	&*&	&	& 6	& &	&	& 		& 0.032	&		& 	& 3\\
12	& 13-Oct-1996	& 50370	& &	&	& 	& 916 &	&	& 		& 14.8	& 9.1		& 23.9 & 6\\
13	& 16-Jan-2000	& 51559	&*& 0.0008 & 0.53 & 3.29 & 19 & 0.02 & 0.31 & 0.193	& 0.236 & 34.5		& 35 & 7\\
14	& 07-Jul-2004	& 53193	&*& 0.0007 & 0.23 & 2.1	& 26.14 & 54 & 0.21 & 0.13	& 0.16	& 22.8		& 77.3 & 8\\
\hline
\multicolumn{14}{l}{[1] \citep{cor88}, [2] \citep{mcc87}, [3] \citep{fla96}, [4] \citep{fla90}, [5] \citep{mcc96} } \\
\multicolumn{14}{l}{[6] \citep{wan00}, [7] \citep{dod02}, [8] \citep{dod07} } \label{tab1}
\end{tabular}
\end{minipage}
\end{table*}

Since 1969, Vela has been seen to glitch a total of 17 times, comprising 15 macro-glitches ($\Delta\nu \sim10^{-6} \nu_{g0}$) and two micro-glitches ($\Delta\nu \sim10^{-8} \nu_{g0}$).  The modified Julian date and timing parameters of each event are listed in Table \ref{tab1}. The first four glitches were discovered in data collected by the Deep Space Network \citep{cor88, rad69,man76}.
Since then, Vela has been monitored almost continuously by two radio telescopes: the 14-\rm{m} antenna at the University of Tasmania's Mt Pleasant Observatory (5 hours per day from 1981 to 1987, 18 hours per day thereafter) \citep{mcc83, mcc87,mcc96,dod02,dod07}, and the 26-\rm{m} antenna at the Hartebeesthoek Radio Observatory (11.5 hours per day since 1984) \citep{fla90, fla96, buc08}.  In their present configurations, the telescopes record all four Stokes parameters at 635/950/1390 \rm{MHz} and 1.644/2.326 \rm{GHz} respectively.  To date they have each detected a total of nine glitches.  Timing parameters have not yet been published for the latest event \citep{buc08}, which does not appear in Table \ref{tab1}.

Continuous timing data is crucial for the work in this paper, because an accurate measurement of the glitch epoch feeds through into accurate measurements of the recovery time-scales $t_n$, the shortest of which can be a fraction of a day \citep{dod02,dod07}.  All glitches for which data were recorded during the event itself are marked with an asterisk in the fourth column of Table \ref{tab1}.  The timing parameters in Table \ref{tab1} are defined in terms of equation (\ref{goeq17}).  Note that the Hartebeesthoek group fits for the derivatives $\Delta\dot{\nu}_n$ rather than $\Delta\nu_n$; the former are converted to the latter in Table \ref{tab1}. Some events were observed simultaneously at Tasmania and Hartebeesthoek, e.g., the 1988 Christmas glitch \citep{fla90}.  As the efforts to continuously monitor Vela have intensified, the timing parameters have been determined ever more accurately. A third, short time-scale $t_3$ has been discernible since 1988, together with a fourth, even shorter time-scale $t_4$ since 2000.  The exceptions are the ``double glitch'' of 1994, where no discernible relaxation was observed in the month between the two events (hence empty entries in Table \ref{tab1}), and the 1996 glitch, which was not observed at either Mt Pleasant or Hartebeesthoek.
Table \ref{tab1} also presents two different sets of timing parameters for the 1988 Christmas glitch.  Both Mt Pleasant and Hartebeesthoek were monitoring continuously at the time and fitted similar values of $\Delta\nu_2$ and $\Delta\nu_3$ but very different values of $\Delta\nu_1$ and $\Delta\nu_p$.  This discrepancy appears to arise from the tail of the recovery, which was tracked by \citet{mcc96} for $707\,$\rm{d} and \citet{fla96} for $96\,$\rm{d}.

It is important to bear in mind that, just because the empirical timing model (\ref{goeq17}) adequately fits the data, it does not mean that this functional form is representative of the underlying physics of the recovery stage.  As more time-scales are revealed by higher resolution data, the possibility grows that the true functional form for the recovery may be something other than a sum of exponentials, even if it is well approximated by the latter.
One such possibility is a ``non-linear'' decay term that ties the intermediate and long-term relaxation to weak and superweak pinning regimes \citep{alp93}.
A second possibility is explored in the remainder of this section, where we show that (i) the simplest hydrodynamic model predicts that $\nu_g(t)$ is not exponential, and (ii) the complete, two-component superfluid model reproduces (\ref{goeq17}) in a natural way for $N\leq3$.

\subsection{One-component viscous fluid} \label{gosec3b}

\begin{figure}
 \includegraphics[width=74mm]{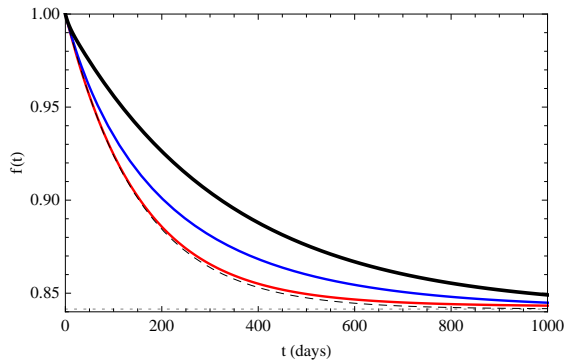}
 \caption{The 1985 Vela glitch, as fitted by a one-component spin-up model (\S\ref{gosec3b}), showing the normalized spin frequency $f(t)$ as a function of time (in \rm{d}).  The data collected by \citet{mcc87} are displayed as a heavy black curve.  The lighter curves correspond to the theoretical model with $K=10^{-2}$ (blue, top) and $K=10^{2}$ (red, bottom).  The dashed curve corresponds to a pure exponential spin down matched to the observed slope at $t=0$; it does not coincide with either the data or the theory. The dotted curve is the steady-state spin frequency $f(\infty)$.}
\label{figvela1}
\end{figure}

We begin by investigating the classical case where there is no superfluid component.  The explicit forms of equations (\ref{goeq10})--(\ref{goeq16}) in this limit are presented in \S\ref{gosecApB}.  The equations involve only one free parameter, $K$.
The classical model is compared to observational data in Figure \ref{figvela1}, using the 1985 glitch as a test case \citep{mcc87}.  The data themselves are plotted as a thick black curve, based on the timing parameters in Table \ref{tab1}.  A pure exponential decay which matches the initial slope and steady state is plotted as a dashed curve.  The remaining curves correspond to $K=10^{-2}$ (blue, top) and $K=10^{2}$ (red, bottom).

Figure \ref{figvela1} demonstrates that a spherical crust experiencing a viscous Ekman torque does not spin down exponentially, even if the spin-up is linear $(\Delta\Omega\ll\Omega)$; the red and blue curves and the dashed curve in Figure \ref{figvela1} do not coincide.
This non-exponential behaviour occurs because the Ekman torque varies with latitude in a sphere.  Hence, even in the simplest model, there is a natural hydrodynamic explanation for multiple time-scales, namely the Ekman time-scale associated with the ``ring'' of crust at each latitude.

Figure \ref{figvela1} demonstrates that there is no value of $K$ which fits the observational data for the 1985 glitch in the simple model.
The same holds true for all the other events in Table \ref{tab1}.
Therefore, we must look to more detailed models to explain the form of $\nu_g(t)$ observed during the recovery stage.

\subsection{Two-component superfluid} \label{gosec3c}

\begin{figure*}
 \includegraphics[width=160mm]{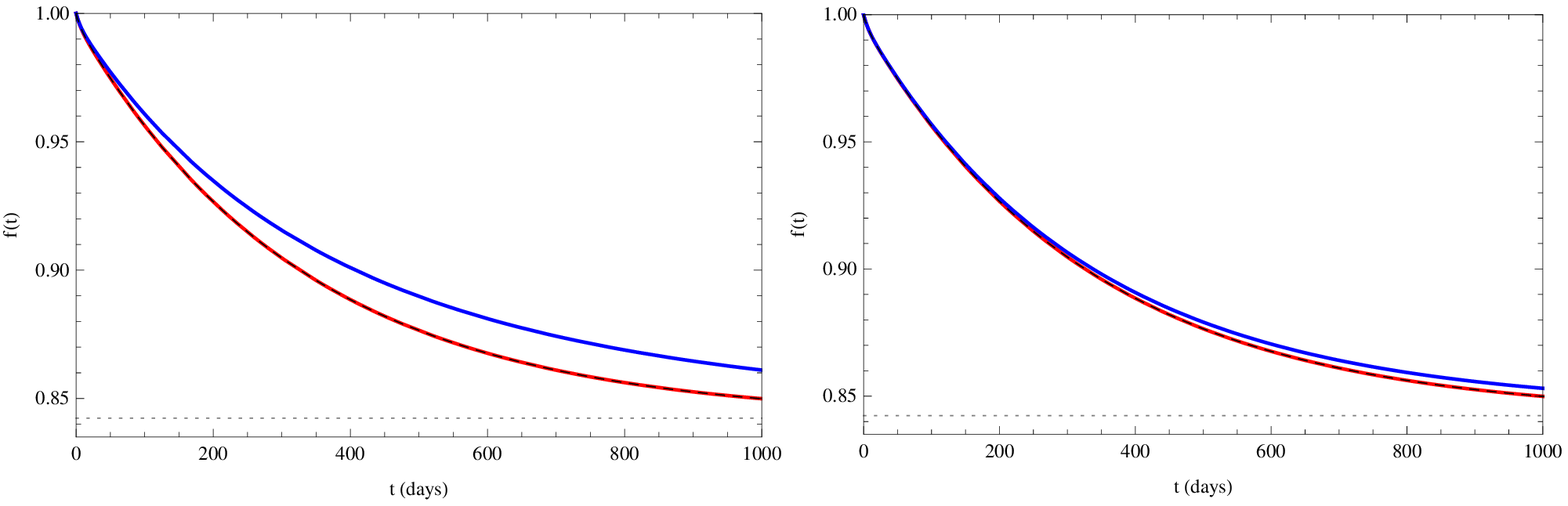}
 \caption{The 1985 Vela glitch, as fitted by a two-component spin-up model (\S\ref{gosec3c}), showing the normalized spin frequency $f(t)$ as a function of time (in \rm{d}).  The blue curve represents $f(t)$ for (a) $B=2.28 \times 10^{-8}$, $E= 9.28\times 10^{-19}$, $\rho_n=0.1$, $K=1$, $\Omega_{n0}=0.38$ and $\Omega_0=0.68$,  and (b) $B=2.52 \times 10^{-8}$, $E=1.16 \times 10^{-21}$, $\rho_n=0.1$, $K=50$, $\Omega_{n0}=0.65$ and $\Omega_0=0.84$.  The \citet{mcc87} data $f_{obs}(t)$ are graphed in red. The approximate solution (\ref{goeq22}) is given by the dashed curve. The dotted curve is the steady-state spin frequency $f(\infty)$.}
\label{figvela85a}
\end{figure*}

Next we apply our complete two-fluid model [equation (\ref{goeq14})] to the 1985 test case. In Figure \ref{figvela85a}, we present two distinct fits to the timing solution measured by \citet{mcc87}.  The model parameters are given in the caption.  The two fits correspond to (a) a heavy crust, with $K=1$ and  $B E^{-1/2}=23.6$, and (b) a light crust, with $ K=50$ and  $B E^{-1/2}=739$ with $\rho_n=0.1$ in both cases.  In both fits, the theoretical curve is drawn in blue and the data is drawn in red.  Also plotted is the approximate solution given by (\ref{goeq22}), which is valid in the regime $E^{1/2}\ll B,B'\ll1$ and $K\gg1$.  We see that (\ref{goeq22}) approximates the exact solution more closely in Figure 2(b), where $B E^{-1/2}$ and $K$ are larger, than in Figure 2(a).

Interestingly, in the regime $E^{1/2}\ll B,B'\ll1$ and $K\gg1$, $f(\tau)$ reduces to the sum of two exponentials.  Yet, for a single viscous fluid, $f(\tau)$ is the convolution of many exponential time-scales at different latitudes (see \S\ref{gosec3b}).  Why, then, does the multi-exponential behavior go away, when the two-component superfluid still contains a substantial viscous component?  The reason is that the mutual friction force and the relatively large fluid inertia dominate the flow dynamics in the above regime.  As a result, the latitudinal variation of time-scales associated with the viscous torque is insignificant in relative terms.


\begin{figure}
 \includegraphics[width=74mm]{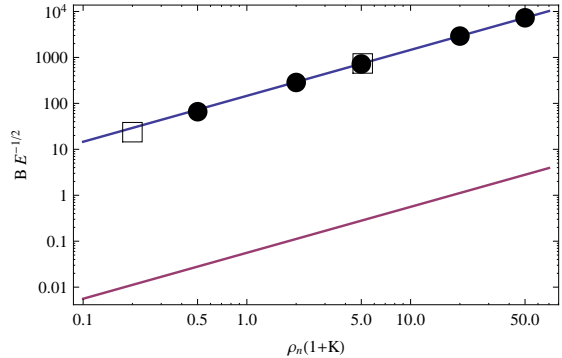}
 \caption{Superfluidity coefficients $B E^{-1/2}$ and $\rho_n (1+K)$ inferred from the 1985 Vela glitch.  The points are obtained by fitting the theoretical solution $f(\tau)$ from (\ref{goeq14}) to the data by eye.   The open boxes correspond to the fits in Figures \ref{figvela85a}(a) and (b).  The filled circles indicate five additional fits.  The two curves are plots of the relation (\ref{goeq26}) obtained by fitting two exponentials to the data, with $(t_1,t_2)=$ (6.5, 332) (top) and (332, 6.5) (bottom) ($t_1$ and $t_2$ in units of \rm{d}).}
\label{figvela85c}
\end{figure}

Figures \ref{figvela85a}(a) and \ref{figvela85a}(b) imply that there is a degeneracy in the fitting parameters for this glitch.
For increasing values of $K$, a collection of accurate fits may be obtained.
Indeed, it transpires that an infinite, one-parameter family of valid fits exists.
This can easily be understood in terms of the approximate solution.
As discussed in \S\ref{gosec2d}, when only two time-scales are resolved in the timing data, $f_{obs}(\tau)$ can be compared directly to (\ref{goeq22}) to yield (\ref{goeq24}) and (\ref{goeq25}).
Combining (\ref{goeq24}) and (\ref{goeq25}), we arrive at the relation
\begin{equation} \label{goeq26}
 B E^{-1/2}=\left( \frac{20 t_2}{7 t_1}\right) \rho_n(1+K) \, ,
\end{equation}
which predicts a direct proportionality between the superfluidity coefficients $B E^{-1/2}$ and $\rho_n (1+K)$, with slope $(20 t_2)/(7 t_1)$.

In Figure \ref{figvela85c}, we plot (\ref{goeq26}) as a straight line on the $B E^{-1/2}$--$\rho_n (1+K)$ plane on a log-log scale.
We also plot the parameters obtained by fitting $f(\tau)$ to the 1985 data by eye, including the fits in Figures 2(a) and 2(b) (open boxes) and five additional fits (filled circles).
The parameters extracted from these fits are quoted in Appendix C.
We find that (\ref{goeq26}) is consistent with the by-eye fits, even in the regime $B E^{-1/2}\sim1$ where the approximate solution nominally breaks down.

The lower line plotted in Figure \ref{figvela85c} is obtained by swapping $t_1$ and $t_2$ in (\ref{goeq24}) and (\ref{goeq25}), as discussed in \S\ref{gosec2d}.  By-eye fits are harder to achieve along this second line, because the iterative numerical solution to (\ref{goeq14}) is unstable.

\begin{figure}
 \includegraphics[width=74mm]{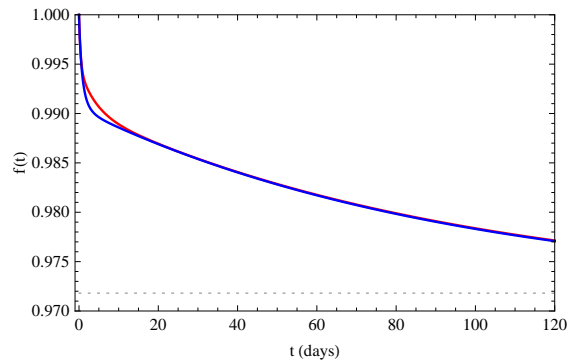}
 \caption{The 1988 Vela glitch, as fitted by a two-component spin-up model \S\ref{gosec3c}, showing the normalized spin frequency $f(t)$ as a function of time (in \rm{d}).  The blue curve represents $f(t)$ for $B=5.0 \times 10^{-9}$, $E=3.05 \times 10^{-15}$, $\rho_n=0.01$, $K=53$, $\Omega_0=\Omega_{n0}=0.97$.  The \citet{fla90} data $f_{obs}(t)$ are graphed in red. The dotted curve is the steady-state spin frequency $f(\infty)$.  The agreement is good.}
\label{figvela88}
\end{figure}

Does the two-component model perform equally well for glitches with three time-scales?  In Figure \ref{figvela88}, we present a fit to the timing solution measured by \citet{fla90} for the 1988 Christmas glitch.  The model reproduces the data admirably.  Moreover, the degeneracy expressed by (\ref{goeq26}) is no longer present; there is just one combination of $B E^{-1/2}$ and $\rho_n(1+K)$ which leads to an acceptable fit.  It is a testament to the power of the model that (i) it handles the triple-time-scale situation without introducing extra parameters, and (ii) the resulting fit constrains the superfluidity coefficients uniquely.

\subsection{Superfluidity coefficients} \label{gosec3d}

\begin{figure}
 \includegraphics[width=74mm]{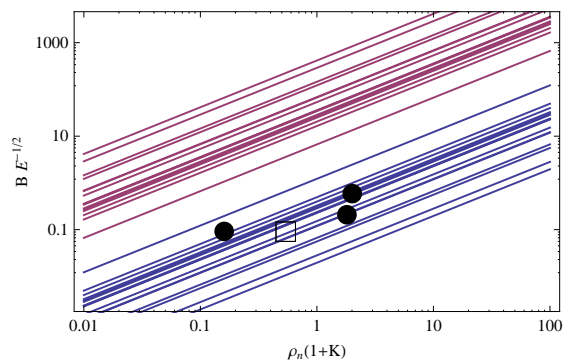}
 \caption{Superfluidity coefficients $B E^{-1/2}$ and $\rho_n (1+K)$ for all Vela glitches with $N\geq2$. The bottom (blue) and top (red) lines correspond to (\ref{goeq26}) as it stands, and with $t_1$ and $t_2$ swapped, respectively. The points correspond to by-eye fits to events with $N\geq3$ (filled circles for Mt Pleasant, open boxes for Hartebeesthoek).}
\label{figvelascat}
\end{figure}

The fitting procedure applied to the 1985 and 1988 test cases in \S\ref{gosec3c} can be extended to all the events in Table \ref{tab1} with $N\geq2$.  In every instance, the two-component model matches the data at least as well as in Figures \ref{figvela85a} and \ref{figvela88}.  The values of $B E^{-1/2}$ and $\rho_n(1+K)$ from each event are plotted in Figure \ref{figvelascat}.  The 16 pairs of diagonal lines come from plotting (\ref{goeq26}) for every glitch with $N\geq2$, taking the time-scales $t_1$ and $t_2$ both ways around; for glitches with $N>2$, the two longest time-scales are assigned to $t_1$ and $t_2$.  The lower lines (blue) correspond to (\ref{goeq26}), and the upper lines (red) correspond to swapping $t_1$ and $t_2$, as discussed in \S\ref{gosec2d}.  Measurements with different telescopes are plotted together.  In addition, there are four events in Table \ref{tab1} (1988, 1991, 2000, 2004) with $N\geq3$, for which the fitting procedure in \S\ref{gosec3c} yields unique values of $B E^{-1/2}$ and $\rho_n(1+K)$.  The four independent measurements of these values are plotted as points in Figure \ref{figvelascat} (open boxes for Mt Pleasant, filled circles for Hartebeesthoek).  In all the fits, the very short ($\sim1$ \rm{min}) time-scale $t_4$ is excluded, as it is probably associated with different (superconducting) physics; see \S\ref{gosec6} and \citet{alp09}.

Given that constitutive properties like viscosity and mutual friction should not vary significantly during the interval between glitches, we expect the values of $B E^{-1/2}$ and $\rho_n(1+K)$ inferred from all the observed Vela glitches to cluster. Figure 5 allows us to test this hypothesis.
We find that the upper lines group around $B E^{-1/2}\rho_n^{-1}(1+K)^{-1}\approx87\pm113$ (mean $\pm$ standard deviation), while the lower lines group around  $B E^{-1/2}\rho_n^{-1}(1+K)^{-1}\approx0.268\pm0.289$.  Moreover, the four glitches with $N\geq3$, for which it is possible to determine $B E^{-1/2}$ and $\rho_n(1+K)$ uniquely, group around $B E^{-1/2}\rho_n^{-1}(1+K)^{-1}\approx 0.80\pm0.90 $.  It is noteworthy that the unique fits favour the ordering $t_1>t_2$; this is consistent with predictions from nuclear theory, as we shall explain in \S\ref{gosec5a}.

For a canonical light crust, with $0.3\leq\rho_n(1+K)\leq3$, the above results imply $26\leq B E^{-1/2}\leq261$ (upper lines) or $0.08\leq B E^{-1/2}\leq0.8$ (lower lines).  In both cases, the mutual friction time-scale is within roughly one order of magnitude of the Ekman time-scale.  We compare the above results with theoretical calculations of $B$ and $E$ in a $^1 S_0$ neutron superfluid and exotic quark matter in \S\ref{gosec5}.

\section{The Crab}  \label{gosec4}

\subsection{Data} \label{gosec4a}

\begin{table*}
 \centering
 \begin{minipage}{1500mm}
  \caption{Timing parameters for large glitches $(\Delta\nu\geq1\rm{\mu Hz})$ in the Crab pulsar $\left(\nu_{g0}=29.9\, {\rm Hz}\right)$.  An asterisk in the CM column denotes continuous monitoring.}
  \begin{tabular}{@{}clrrrrrrrrrrc@{}}
  \hline
   Glitch & Date & MJD & CM &  $t_3$ & $t_2$ & $t_1$ & $\Delta\nu_3$ & $\Delta\nu_2$ & $\Delta\nu_1$ & $\Delta\nu_p$ & $\Delta\nu$ & Ref. \\
    no.& & &  & \multicolumn{3}{c}{{\rm d}} & \multicolumn{5}{c}{$\mu${\rm Hz}} &  \\
 \hline
1	& 30-Sep-1969	& 40494	& &	& 18.7	&  	&	& 0.07	& 	& 0.05	& 0.12 	& 1 \\
2	& 04-Feb-1975	& 42448	& &	& 18	& 97 	&	& 1.01	&$-$0.71& 1.02	& 1.32 	& 1 \\
3	& ??-???-1981	&~44900	& &	& 	& 222	&	& 	&$-$0.28& 	&	& 1 \\
4	& 22-Aug-1986	& 46664	&*&	& 9.3	& 123 	&	& 0.12	&$-$0.11& 0.11	& 0.12	& 1 \\
5	& 29-Aug-1989	& 47767	&*& 0.8	& 18	& 265 	&$-$0.7	& 2.28	&$-$2.11& 2.38 	& 1.85	& 1 \\
6	& 21-Nov-1992	& 48947	& &	& 2	&  	& 	& 0.26	& 	& 0.4	& 0.3	& 1 \\
7	& 30-Oct-1995	& 50021	& &	& 3.2	&  	& 	& 0.064	& 	& 0.015	& 0.08	& 1 \\
8	& 25-Jun-1996	& 50260	& & 0.5	& 10.3	&  	&$-$0.31& 0.66	& 	& 0.31	& 0.66	& 1 \\
9	& 11-Jan-1997	& 50459	& &	& 3.0	&  	& 	& 0.2	& 	& 0.032	& 0.23	& 1 \\
10	& 10-Feb-1996	& 50489	& &	& 2.2	&  	& 	&$-$0.03& 	& 0.05	& 0.02	& 1 \\
11	& 30-Dec-1996	& 50813	& &	& 2.9	&  	& 	& 0.24	& 	& 0.017	& 0.26	& 1 \\
12	& 01-Oct-1999	& 51452	& &	& 3.4	&  	& 	& 0.24	& 	& 0.04	& 0.29	& 1 \\
13	& 16-Jul-2000	& 51741	& &	& 4.0	&  	& 	& 0.584	& 	& 0.143	& 0.73	& 2 \\

\hline
\multicolumn{13}{l}{[1] \citep{won01}, [2] \citep{wan01} }  \label{tab2}
\end{tabular}
\end{minipage}
\end{table*}

Since its discovery in 1968, the Crab pulsar has been monitored daily for glitch activity by a number of groups \citep{gro75,gul77,loh81}.  The most extensive and ongoing effort is overseen by the Jodrell Bank Observatory, using a 12.5-\rm{m} dish to measure all four Stokes parameters at 610 \rm{MHz} for a maximum of 14 hr a day, occasionally supplemented by readings at 1.4 \rm{GHz} using the 76m Lovell telescope \citep{lyn93,she96,lyn00}.  A total of 26 glitches have been observed \citep{mel08}.  Of these, 13 have been monitored sufficiently regularly to resolve the recovery stage.  The timing parameters of the latter events are collected in Table \ref{tab2}.
Although the 1969 and 1975 glitches were observed by several telescopes, Jodrell Bank was the only group to fit the data to the timing model given by (\ref{goeq17}).  Table \ref{tab2} makes it clear that it is harder to resolve multiple, distinct, exponential decays in the Crab than in Vela.  Only the 1989 glitch has $N=3$ \citep{lyn92}, and this event (the famous ``slow glitch'') is anomalous anyway, because the timing solution describes the spin-up event as well as the recovery stage, the only time the spin up has been resolved.  As the spin up involves vortex physics outside the scope of the models in this paper, we exclude the 1989 glitch from our analysis.  For all the other events, we have $N\leq2$.

Crab glitches differ from Vela glitches in a number of respects.  The waiting-time probability distribution function is fitted accurately by an exponential in the Crab, i.e. it is consistent with a Poisson process \citep{mel08}.  By contrast, Vela glitches quasi-periodically, with a small subset ($\sim20\%$) of events spaced roughly evenly \citep{mel08}.  The Crab glitches more frequently, at an average rate of $0.91^{+0.4}_{-0.3}\, \rm{yr^{-1}}$, compared to $0.43^{+0.19}_{-0.16}\, \rm{yr^{-1}}$ for Vela \citep{mel08}. Its frequency jumps $\Delta\nu$ are measured to be $\sim1\%$ of Vela's.  During the recovery stage, the greatest distinction is the ``overshoot'' observed in the Crab, wherein the crust decelerates below its steady-state angular velocity before rising again asymptotically.  This is characterized by the negative values of $\Delta\nu_n$ in Table \ref{tab2}, which do not occur for Vela.
The Crab has not attracted the same level of continuous monitoring as Vela, so Table \ref{tab2} is sparser than Table \ref{tab1}. More experiments are urgently required to investigate the form of the overshoot in more detail.

\subsection{Two-component superfluid} \label{gosec4b}

\begin{figure*}
 \includegraphics[width=150mm]{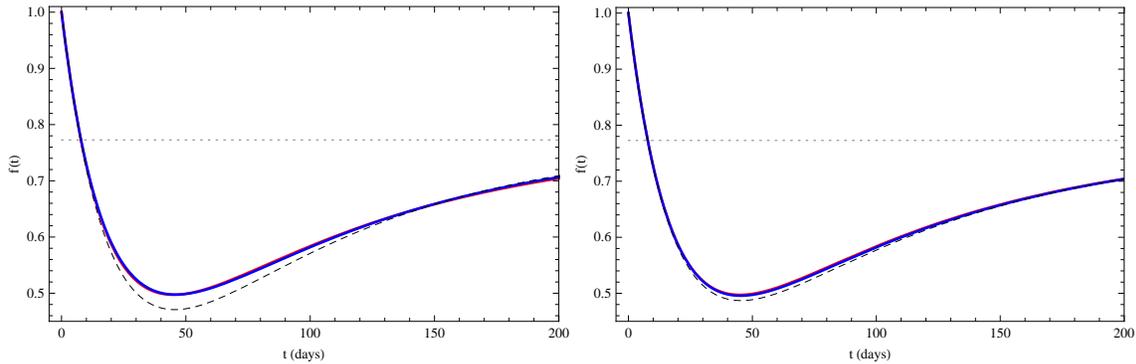}
 \caption{The 1975 Crab glitch, as fitted by a two-component spin-up model (see \S\ref{gosec3c}), showing the normalized spin frequency $f(t)$ as a function of time (in \rm{d}).  The blue curve represents $f(t)$ for (a) $B=7.3 \times 10^{-10}$, $E=1.83 \times 10^{-21}$, $\rho_n=0.01$, $K=2500$, $\Omega_{n0}=0.25$ and $\Omega_0=0.77$,  and (b) $B=3.3 \times 10^{-9}$, $E=8.44 \times 10^{-23}$, $\rho_n=0.01$, $K=2500$, $\Omega_{n0}=-2.47$ and $\Omega_0=0.77$.  The \citet{lyn93} data $f_{obs}(t)$ are graphed in red. The approximate solution (\ref{goeq22}) is given by the dashed curve. The dotted curve is the steady-state spin frequency $f(\infty)$.}
\label{figcrab75}
\end{figure*}

We repeat the analysis in \S\ref{gosec3c} using the data in Table \ref{tab2}.  To begin with, we apply the complete two-fluid model [equation (\ref{goeq14})] to the 1975 glitch as a test case.  In Figure \ref{figcrab75} we present two distinct fits to the timing solution measured by \citet{lyn93}.  Again, we find that the approximate solution (dashed curve) adheres more closely to the analytical solution (blue curve) in panel (b), where $B E^{-1/2}=3.6\times10^{2}$ is larger than in panel (a) ($B E^{-1/2}=17$).  In both panels, we have $\rho_n=0.01$ and $K=2500$, demonstrating that for $\rho_n(1+K)=25$ there are two possible fits.  These correspond to swapping the time scales $t_1$ and $t_2$, as discussed in \S\ref{gosec3c}.

\begin{figure}
 \includegraphics[width=74mm]{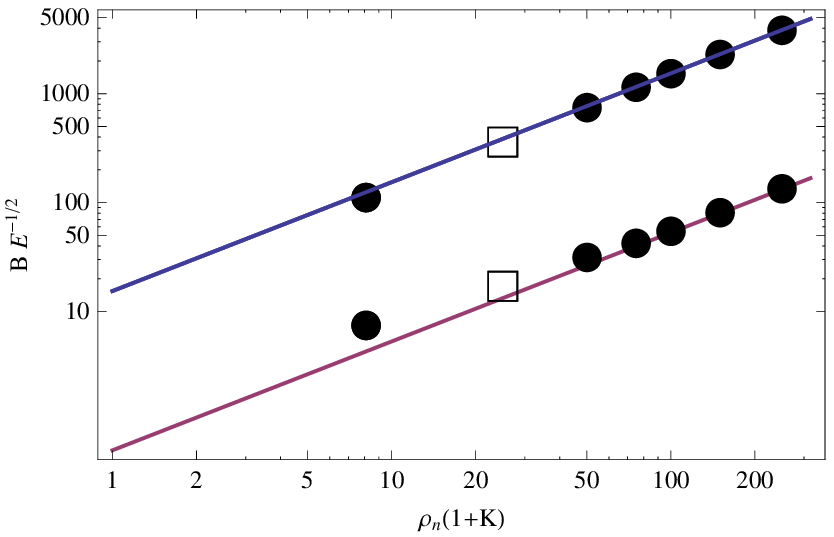}
 \caption{Superfluidity coefficients $B E^{-1/2}$ and $\rho_n (1+K)$ inferred from the 1975 Crab glitch.  The points are obtained by fitting the theoretical solution $f(\tau)$ from (\ref{goeq14}) to the data by eye.   The open boxes correspond to the fits in Figures \ref{figcrab75}(a) and (b).  The filled circles indicate 12 additional fits.  The two curves are plots of the relation (\ref{goeq26}) obtained by fitting two exponentials to the data, with $(t_1,t_2)=$ (18, 97) (top) and (97, 18) (bottom) ($t_1$ and $t_2$ in units of \rm{d}).}
\label{figcrab75b}
\end{figure}

Next, in Figure \ref{figcrab75b}, we plot a range of fits obtained for the 1975 glitch on the $B E^{-1/2}$--$\rho_n (1+K)$ plane, c.f., Figure \ref{figvela85c} for Vela.  The parameters inferred by fitting the 1975 glitch by eye are marked by solid circles, and the data from Figure \ref{figcrab75} are marked by open boxes.  Again, we find that (\ref{goeq26}) matches admirably the by-eye fits when there are two time-scales, even in the regime where the approximate solution nominally breaks down.  Indeed, the overall agreement is better than for Vela because the iterative numerical solution to (\ref{goeq14}) is stable on the bottom branch too, cf. Figure \ref{figvela85c}.

\subsection{Overshoot} \label{gosec4c}
Figure \ref{figcrab75} makes it plain that the analytical model captures the ``overshoot'' in the Crab's recovery naturally. We can exploit the approximate solution (\ref{goeq22}) to pin down the physical conditions required for the overshoot to occur.  As noted above, an overshoot requires $\Delta \nu_n < 0 $ for one value of $n$.  In the context of equation (\ref{goeq22}), this translates into
\begin{equation} \label{goeq27}
 C=\frac{\Omega_0-\Omega_{n0}}{7 B E^{-1/2}/(20 \rho_n K)-1}<0\, ,
\end{equation}
or else
\begin{equation} \label{goeq28}
 1-C-f(\infty)=\frac{7 B E^{-1/2}(1-\Omega_0)/(20 \rho_n K) +\Omega_{n0}-1}{7 B E^{-1/2}/(20 \rho_n K)-1}<0\, .
\end{equation}
We can now distinguish four cases, depending on the sign of the denominator and whether (\ref{goeq27}) or (\ref{goeq28}) is satisfied.   If the denominator is negative, i.e. for $(7 B E^{-1/2})/(20 \rho_n K)<1$, and (\ref{goeq28}) is satisfied, we must have $1<\Omega_0<\Omega_{n0}$.  However, this ordering produces a different type of overshoot, where the angular velocity of the crust initially increases above its steady-state value before decreasing asymptotically. On the other hand, if (\ref{goeq27}) is satisfied, the condition for an overshoot becomes $\Omega_{n0}<\Omega_0<1$.  This is the case in Figure \ref{figcrab75}(a), where we have $(7 B E^{-1/2})/(20 \rho_n K)=0.186$, $\Omega_{n0}=0.25$ and $\Omega_0=0.77$.  Similarly, if the denominator is positive, i.e. for $(7 B E^{-1/2})/(20 \rho_n K)>1$, then we must have  $1<\Omega_0<\Omega_{n0}$ if (\ref{goeq27}) is satisfied (which is not the overshoot we are looking for) and $\Omega_{n0}<\Omega_0<1$ if (\ref{goeq28}) is satisfied.  The latter is the situation in Figure \ref{figcrab75}(b), where we have $(7 B E^{-1/2})/(20 \rho_n K)=5.39$, $\Omega_{n0}=-2.47$ and $\Omega_0=0.77$.  Therefore the condition for an overshoot of the form observed in the Crab is always
\begin{equation} \label{goeq29}
 \Omega_{n0}<\Omega_0<1 \, .
\end{equation}

We can understand (\ref{goeq29}) physically in the context of Figure \ref{figcrab75}.   In panel (a), the crust is light and it initially decelerates rapidly below its steady-state angular velocity, in response to the viscous torque. Then, over a longer time-scale, mutual friction spins up the crust and viscous component into corotation with the inviscid component.  In panel (b), the initial differential rotation between the viscous and inviscid components is rapidly removed by mutual friction.  While this is occurring, the even larger initial differential rotation between the viscous fluid and the crust drags the latter below its steady-state angular velocity.  The viscous and invicid components achieve corotation first, followed by the crust, which is brought into corotation over the longer, viscous time-scale.  Of these two possibilities, (a) appears to be the most natural; in the case of (b) a large and negative $\Omega_{n0}$ is required to generate an overshoot.

The condition (\ref{goeq29}) for an overshoot lends some insight into the glitch trigger. Firstly, glitches leading to an overshoot can occur for all values of $B E^{-1/2} \rho_n^{-1} (K+1)^{-1}$ in principle.  Secondly, the internal fluid components must satisfy (\ref{goeq29}) at $t=0+$, i.e., immediately after the impulsive spin-up of the crust.  The fact that overshoots are observed in the Crab suggests that (\ref{goeq29}) is satisfied for the majority of its glitches, whereas (\ref{goeq29}) is never satisfied in Vela.  These differences must arise from the microphysics in the two objects, such as the distribution of vortex pinning strengths which is plausibly a function of age [e.g. if lattice defects anneal over time \citep{mel09}].



\subsection{Superfluidity coefficients} \label{gosec4d}

\begin{figure}
 \includegraphics[width=74mm]{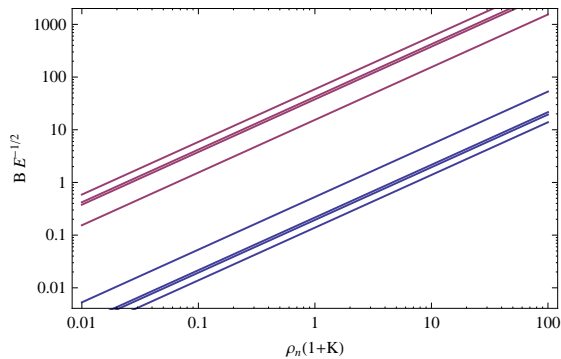}
 \caption{Superfluidity coefficients $B E^{-1/2}$ and $\rho_n (1+K)$ for all Crab glitches with $N\geq2$. The bottom (blue) and top (red) lines correspond to (\ref{goeq26}) as it stands, and with $t_1$ and $t_2$ swapped respectively.}
\label{figcrabscat}
\end{figure}
Finally, in Figure \ref{figcrabscat}, we collect the inferred parameters for all Crab glitches with $N\geq2$ on the $B E^{-1/2}$--$\rho_n (1+K)$ plane, c.f., Figure \ref{figvelascat} for Vela.
The four pairs of diagonal lines come from plotting (\ref{goeq26}) for every glitch with $N=2$, taking the time-scales $t_1$ and $t_2$ both ways around.
The lower lines (blue) correspond to (\ref{goeq26}) as it stands, while the upper lines (red) correspond to swapping $t_1$ and $t_2$ in (\ref{goeq26}).  We find $B E^{-1/2}\rho_n^{-1}(1+K)^{-1}\approx38.5\pm17.9$ (mean $\pm$ standard deviation) for the upper lines, and $B E^{-1/2}\rho_n^{-1}(1+K)^{-1}\approx0.270\pm0.177$ for the lower lines.  For a canonical light crust, with $0.3\leq\rho_n(1+K)\leq3$, we therefore have $12\leq B E^{-1/2}\leq116$ (upper lines) or $0.08\leq B E^{-1/2}\leq0.81$ (lower lines).

Encouragingly, the Crab results are similar to Vela, lending general support to the model.
The concordance is especially significant at $t_1$ and $t_2$ differ significantly in the two objects.
In the upper fits, the mean value of $B E^{-1/2}\rho_n^{-1}(1+K)^{-1}$ for the Crab is about half that of Vela, while the mean in both pulsars is roughly the same for the lower fits in both objects.
In general,  typical values of $B$ and $E$ individually are slightly lower in the Crab than in Vela.
For example, upon comparing Figure \ref{figcrab75} with Figure \ref{figvela85a}, we find $10^{-10}\la B \la 10^{-9}$ and $10^{-23} \la E \la 10^{-19}$ in the Crab, as against  $B\approx10^{-8}$ and $10^{-23}\la E \la 10^{-21}$ for Vela.
In \S\ref{gosec5}, we discuss these numbers in the context of recent theoretical and experimental constraints obtained from nuclear theory and heavy-ion collider experiments.

\subsection{Magnetically coupled crust and core superfluid} \label{gosec4e}

\begin{figure}
 \includegraphics[width=74mm]{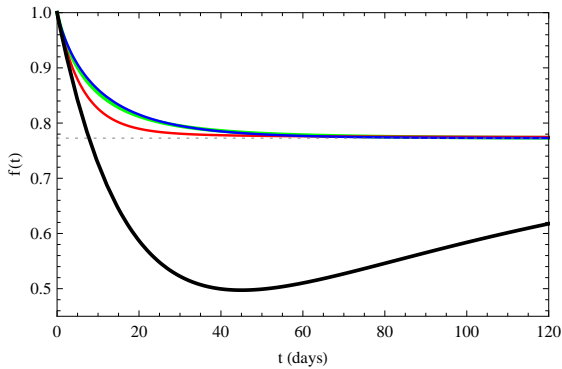}
 \caption{The 1975 Crab glitch, as fitted by a locked two-component spin-up model with $B\sim1$ and $\rho_n=0.01$, showing the normalized spin frequency $f(t)$ as a function of time (in \rm{d}).  The observational data $f_{obs}(t)$ are displayed as a heavy black curve.  The lighter curves correspond to the theoretical model with $K=10^{-2}$ (blue, top), $K=1$ (green, centre) and $K=10^{2}$ (red, bottom). The dotted curve is the steady-state spin frequency $f(\infty)$.}
 \label{figure11}
\end{figure}

The hydrodynamic model presented here can be used to represent and test a variety of neutron star pictures.
One important possibility, discussed in \S\ref{gosec1}, \S\ref{gosec2a} and \S\ref{gosec2b}, is that the viscous and inviscid fluid components lock together via magnetic forces (e.g. if the proton-electron plasma is strongly coupled to the neutron superfluid via entrainment).
This regime corresponds to $B\sim1$ in our model (see \S\ref{gosec2b}).
Under such conditions, equation (\ref{goeqa01}) for the azimuthal velocity component of the interior flow  reduces to
\begin{equation} \label{goeq29a}
 v_\phi=-r \omega_+ \int_0^\tau d\tau'e^{\omega_+\left(\tau-\tau'\right)}f\left(\tau'\right)+r\Omega_0 e^{\omega_+\tau}
\end{equation}
with
\begin{eqnarray}
 \omega_+&=&-\frac{\rho_n J(r)}{h(r)} \label{goeq29b} \\
 \omega_-&=&0 \label{goeq29c}
\end{eqnarray}
Importantly, because of (\ref{goeq29c}), there is now only a single time-scale in (\ref{goeq29a}), which resembles the equation for a single-component fluid [see \S\ref{gosec3b} and (\ref{goeqb01})].

In Figure \ref{figure11}, we attempt to fit the 1975 Crab glitch using (\ref{goeq13}) and (\ref{goeq29a}).
As we can see, for $10^{-2}<K<10^2$, the theory produces a nearly exponential decay on a single time-scale.
Therefore, strong coupling $(B\sim1)$ de-activates the time-scale on which the viscous and inviscid  components come into co-rotation, effectively reducing the problem to a single fluid.
As in \S\ref{gosec3b}, this limiting case does not contain enough freedom to fit the observational data.
{\it Most importantly, we find that it never leads to an overshoot. }
Therefore, to reproduce the glitch recovery in the Crab, we must leave both time-scales in the problem, with their ratio to be determined by observations.

Similar arguments can be made for the coupling of the viscous fluid (proton-electron plasma) to the crust.
Theory predicts this coupling to be strong ($\sim1\,\rm{s}$) if the spin up is a result of induced tension in the magnetic field lines, or $\sim30\,\rm{s}$ if it is a result of classic Ekman pumping \citep{eas79}.
This scenario corresponds to $K\sim1$ in our model.
If the only remaining component is the neutron superfluid, it then responds on the mutual friction time-scale  $(B \Omega)^{-1}$.
Again, only one remaining time-scale remains with which to fit the data.

We emphasize that the failure of the above strongly coupled regime to fit the data does {\it not} mean that the neutron condensate and the proton-electron plasma are weakly coupled.
The star might consist of multiple superfluid components [e.g. multiple moments of inertia \citep{alp93,alp96}] or a significant uncharged inviscid component (non-ideal excitations like in helium II, see \S\ref{gosec2}).
Such scenarios are contained in our model by taking $K\sim 1$.

\section{Dissipative processes in bulk nuclear matter} \label{gosec5}

The transport coefficients of bulk nuclear matter at $\sim\,$\rm{MeV} energies have not yet been measured in terrestrial experiments.
Of the 19 coefficients identified by \citet{and06b} in their flux-conservative formalism, only the shear viscosity $\eta$ has been measured in relativistic heavy-ion colliders \citep{adl03,ada07}.
These experiments ($\sim500$ nucleons at $\sim10^2$\rm{GeV}) are conducted far from the physical conditions in a neutron star ($\sim10^{57}$ nucleons at $\sim1\,\rm{MeV}$).
Even so, they have thrown up the interesting finding that $\eta$ approaches the quantum lower bound $\eta/s=\hbar/4\pi k_B$ (where $s$ is the specific entropy) inferred from the duality between anti-de-Sitter and conformal field theories \citep{adl03}.

In this section, we use the theory in \S\ref{gosec2} and the fits to the Vela and Crab data in \S\ref{gosec3} and \S\ref{gosec4} to constrain two transport coefficients:  the shear viscosity $\eta$, and the mutual friction coefficient $B$.  We also constrain the density ratio of the viscous and inviscid components, and hence, indirectly, the superfluidity transition temperature.  As the idealized model in \S\ref{gosec2} does not incorporate stratification, the data are interpreted in terms of uniform $\eta$ and $B$, effectively representing mass-weighted averages of the depth-dependent $\eta(r)$ and $B(r)$ in the real star.  The reader is cautioned to bear this in mind when interpreting the findings.  For example, if $\eta$ and $B$ are very different in the inner and outer core, the mass-weighted average may be a poor approximation to both regions.


\subsection{Neutron-rich matter} \label{gosec5a}

In the outer core [$1.6\times 10^{14}\leq \rho/\rm{gcm^{-3}} \leq 3.9 \times 10^{14}$], the neutrons condense into a $^1S_0$ superfluid.  In this phase, it is surmised that $\eta$ arises from electron-electron scattering outside the superfluid vortices \citep{cut87}, while $B$ arises from electrons scattering off vortex cores \citep{alp84, men91b}.  The bulk transport coefficients can be related to microscopic quantities, like the charged-fluid-neutron-vortex relaxation time and Kelvin-wave oscillation frequency, in the low-frequency, long-wavelength limit by generalising the method of \citet{hal56b}.  \citet{men91b} calculated theoretically that one has
\begin{equation} \label{goeq30}
 \mu=6.0\times10^{20} \rm{g\,cm^{-1} s^{-1} }\left(\frac{\rho}{10^{14} \rm{g\,cm^{-3}}}\right)^2\left(\frac{\it{T}}{10^7 \rm{K}}\right)^{-2}\,,
\end{equation}
\begin{equation} \label{goeq31}
 B=1.1\times10^{-2}\frac{\left(y-1\right)^2 x^{7/6}}{y^{1/2}\left(1-x\right)} \left(\frac{\rho}{10^{14} \rm{g\, cm^{-3}}}\right)^{1/6}\,,
\end{equation}
\begin{equation} \label{goeq32}
 B'=B^2\,,
\end{equation}
where $x$ is the proton fraction $\rho_n/\rho$, and $0.3\leq y = m_p^*/m_p \leq0.7$ is the normalized effective mass of the proton from Fermi liquid theory.  For a cold equation of state below neutron drip, $x$ is given by Eqn. (2.5.16) in \citet{sha83}, viz.
\begin{equation} \label{goeq33}
 \frac{x}{1-x}=\frac{1}{8}\left(1+\frac{1}{x_s^2}\right)^{-1}\left[1+\frac{4Q}{m_n x_s^2}+\frac{4\left(Q^2-m_e^2\right)}{m_n^2 x_s^4}\right]^{3/2}\, ,
\end{equation}
with
\begin{equation} \label{goeq34}
 x_s=\left(\frac{\rho_s}{6.1\times 10^{15} \rm{g\,cm^{-3}}}\right)^{1/3}\, ,
\end{equation}
$Q=m_n-m_p$, and hence $2.6\times 10^{-3}\leq x/(1-x)\leq0.125$; the lower bound occurs at $\rho_s=7.8\times10^{11}\rm{g\,cm^{-3}}$.
Clearly, $\eta$, $B$, and $B'$ are functions of depth through $\rho$, $T$, and hence $x$ (and $m_p^*$, weakly).

\begin{figure}
 \includegraphics[width=74mm]{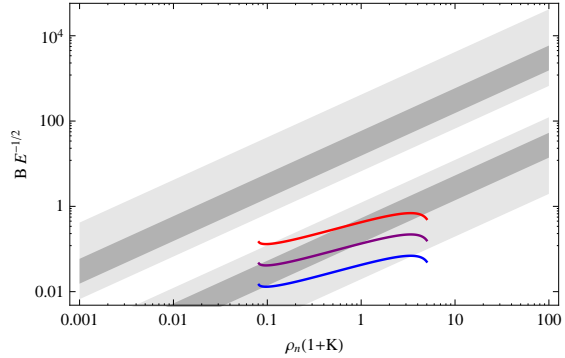}
 \caption{Ranges of the superfluidity coefficients $B E^{-1/2}$ and $\rho_n (1+K)$ predicted by nuclear theory [equations (\ref{goeq30})--(\ref{goeq34})] and inferred from pulsar observations (Figures \ref{figvelascat} and \ref{figcrabscat}).
The lightly and darkly shaded regions correspond to Vela and Crab data respectively.  The theoretical curves are obtained by varying the density $\rho$ for three different values of temperature, $T=10^{7.5}\rm{K}$  (top, red), $T=10^{7.0}\rm{K}$  (centre, green) and $T=10^{6.5}\rm{K}$  (bottom, blue), with $K=50$.  }
\label{fig9}
\end{figure}

\begin{figure}
 \includegraphics[width=74mm]{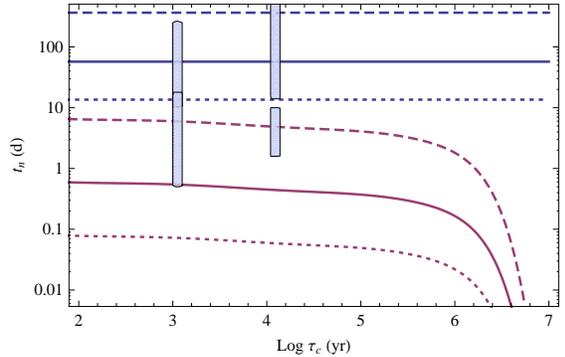}
 \caption{Theoretical mutual friction and viscous time-scales $t_1$ (blue, upper) and $t_2<t_1$ (red, lower), given by (\ref{goeq24}) and (\ref{goeq25}) respectively, as functions of pulsar spin-down age $\tau_c$ (in \rm{yr}). Input quantities $E, \rho_n$ and $B$ are determined from nuclear theory by (\ref{goeq30})--(\ref{goeq34}), assuming standard cooling to get a $T$--$\tau_c$ relation. Curves correspond to densities $\rho=10^{14}$ (dashed curve), $10^{15}$ (solid curve), and $10^{16}$ \rm{g cm$^{-3}$} (dotted curve).  The vertical bars denote the ranges of $t_1$ and $t_2$  measured observationally for the Crab (left) and Vela (right). }
\label{fig10}
\end{figure}

In Figure \ref{fig9}, we consolidate the ranges for $B E^{-1/2}$ and $\rho_n (1+K)$ for Vela (Figure \ref{figvelascat}) and the Crab (Figure \ref{figcrabscat}) onto one diagram.
The lightly shaded regions in Figure \ref{fig9} cover the range of Vela fits presented in Figure \ref{figvelascat}.
In the lower region, $t_1$ and $t_2$ are given by (\ref{goeq24}) and (\ref{goeq25}) respectively; in the upper region, $t_1$ and $t_2$ are swapped.
Similarly, the darkly shaded regions cover the range of Crab fits presented in Figure \ref{figcrabscat}.
Overplotted are three theoretical curves for $B E^{-1/2}$ versus $\rho_n (1+K)$, constructed from equations (\ref{goeq30})--(\ref{goeq34}).
To plot these curves, we assume a canonical fluid-crust ratio, $K=50$.  (It is easy to slide the curves from left to right as $K$ increases.)
The three curves correspond to three different values of temperature, $T=10^{7.5} \rm{K}$  (top, red), $T=10^{7.0} \rm{K}$  (centre, purple) and $T=10^{6.5} \rm{K}$  (bottom, blue).
The density $\rho_s\approx\rho$  increases along each curve, from $\rho=10^{12}$ \rm{gcm$^{-3}$} at the left-hand vertex, to $\rho=10^{17}$ \rm{gcm$^{-3}$} at the right-hand vertex.

Figure \ref{fig9} contains several important lessons.
First and foremost, the three theoretical curves predicted by standard nuclear theory fall naturally on top of the parameter ranges inferred from the Crab and Vela data.
This is strong circumstantial evidence in favor of the spin-down model in \S\ref{gosec2} and \citet{van10}.
It motivates further testing of the model as fresh data become available in the future.
Second, the theoretical curves overlap with the shaded bands in the regime $B E^{-1/2}\rho_n^{-1} (1+K)^{-1}<1$, where the viscous time-scale $t_2$ given by (\ref{goeq25}) is less than the mutual friction time-scale $t_1$ given by (\ref{goeq24}) \citep{men91b, rei93}.
Interestingly, the by-eye fits to the Vela glitches with $N=3$ (points in Figure \ref{figvelascat}) also favour the $t_1>t_2$ regime.
Third, if we turn the argument around and treat $\eta$ and $B$ as known and given by (\ref{goeq30})--(\ref{goeq34}), then the intersection between the theory and data constrains the fluid-crust ratio to $0.08\la\rho_n(1+K)\la4$.
The latter range is independently consistent with the standard nuclear equation of state.
We discuss this issue further in \S\ref{gosec5c}.

The evolution of the recovery time-scales $t_1$ and $t_2$ with pulsar spin-down age $\tau_c$ can also be predicted from (\ref{goeq30})--(\ref{goeq34}), given a relation between $T$ and $\tau_c$.
We assume standard cooling via the Urca and modified Urca processes and a two-zone, heat blanket model \citep{pag98, mel07,per07}.
In Figure \ref{fig10}, we plot $t_1$ and $t_2$, defined by (\ref{goeq24}) and (\ref{goeq25}), versus $\tau_c$, using the theoretical formulas for viscosity and mutual friction in (\ref{goeq30})--(\ref{goeq34}) and taking $t_1>t_2$, in keeping with the conclusions of the previous paragraph.
Results for three different values of $\rho$ are presented; $\rho=10^{14}\,\rm{g\,cm^{-3}}$ (dashed curve), $\rho=10^{15}\, \rm{g\,cm^{-3}}$ (solid curve) and $\rho=10^{16} \rm{g\,cm^{-3}}$ (dotted curve).
The upper (blue) curves for $t_1$ are flat (assuming $m_p^*/m_p$ depends weakly on $T$).
The lower (red) curves indicate that $t_2$ decreases as $T$ drops and $\eta$ decreases.
Overplotted are the observed ranges of $t_1$ and $t_2$ for Vela and the Crab, given in Tables \ref{tab1} and \ref{tab2} respectively.
We see that the theoretical curves are consistent with the observations for $10^{14}\, \rm{g\,cm^{-3}}<\rho<10^{15}\, \rm{g\,cm^{-3}}$, nicely bracketing the expected density of the outer core \citep{per05}.
The predicted decline of $t_2$ for $\tau_c\geq10^6 \rm{yr}$, where neutrino cooling gives way to photon cooling, is an interesting test of the spin-down model.
Glitches observed in pulsars older than $\sim10^6\rm{yr}$ are predicted to recover rapidly ($\la 1\,{\rm d}$) immediately after the glitch (due to increased viscosity), followed by a slower recovery ($\ga 10\,{\rm d}$) mediated by mutual friction.

\subsection{Strange quark matter} \label{gosec5b}

The analysis in \S\ref{gosec5a} can be repeated for the numerous exotic fluid phases involving strange quark matter, which have been hypothesised to exist in neutron star interiors.  Some of these phases are confined to the inner core, where the Ekman process may not penetrate due to stratification \citep{abn96, van09}.  Nevertheless, the analysis in this paper can distinguish between these phases in principle, by extracting values of $\eta$ and $B$ from glitch recovery data.  It is therefore an aid in testing conjectures regarding the susceptibility of quark stars to r-mode instabilities \citep{mad00,man09} and starquake glitches \citep{man07}.

In its simplest manifestation, deconfined strange quark matter is ungapped, i.e., diquark pairing does not occur.  Under these circumstances, the shear viscosity arises predominantly from quark-quark scattering (modified by Landau damping) and is given by \citep{hei93,jai08, sad08,sht08}
\begin{equation} \label{goeq35}
 \eta=5\times10^{15}\left(\frac{\alpha_s}{0.1}\right)^{-3/2}\left(\frac{n}{n_0}\right)^{14/9}\left(\frac{T}{10^9 {\rm K}}\right)^{-5/3} \rm{g\,cm^{-1}s^{-1}}\,.
\end{equation}
In (\ref{goeq35}), $\alpha_s$ denotes the QCD coupling constant and $n_0=0.15\, {\rm fm^{-3}}$ is the nuclear saturation density (with $\rho / m_n \approx 5\times n_0$ typically).  For densities and temperatures pertinent to the core, e.g. $10^{14}\, {\rm gcm^{-3}} < \rho < 10^{15}\,{\rm gcm^{-3}}$ and $10^{8.5}\,{\rm K}<T<10^{9.5}\,{\rm K}$, equation (\ref{goeq35}) predicts $4 \times 10^{-15}<E<6 \times 10^{-13}$.

If diquark pairing does occur, the shear viscosity is suppressed exponentially by a factor ${\rm exp}(-\Delta/3 k_B T)$, where $\Delta$ is the pairing energy gap \citep{mad00}.  For $\Delta\ga 1\,{\rm MeV}$, the suppression is extreme.  Several condensed phases have been identified, two of which we consider here: the color-flavor locked (CFL) and two-flavor superconducting (2SC) phases.  In the CFL phase ($n \la 10 n_0 $, $T\la 0.1\rm{MeV}$),  $u$, $d$ and $s$ quarks pair up to form a condensate that is antisymmetric in its color and flavor indices \citep{alf99, alf01}. Under these conditions, $\eta$ is dominated by phonon-phonon and kaon-kaon scattering \citep{mad00, jai08, alf09,man09}; for $T\la 0.1 \rm{MeV}$, the electron population is exponentially suppressed and scatters proportionately weakly \citep{raj01}.  For the phonon-phonon process (ignoring vortex scattering and ``dressed photons''), the shear viscosity is nominally given by \citep{man05, jai08}
\begin{equation} \label{goeq36}
 \eta=3.7\times 10^6 \left(\frac{\mu_q}{\rm{MeV}}\right)^8\left(\frac{T}{10^9 \rm{K}}\right)^{-5} \rm{g cm^{-1} s^{-1}}
\end{equation}
where $\mu_q\approx300\,\rm{MeV}$ is the quark chemical potential.  For temperatures and densities pertinent to the core, we obtain $5 \times 10^{-11} < E < 5 \times 10^{-5}$.
Superficially, this phase appears to be ruled out by the consolidated data in Figure \ref{fig9}, but it is vital to note that the phonon-phonon mean free path exceeds the stellar radius for $T\la 10^{10}$ K, so the above value of $\eta$ is only relevant for very hot cores, e.g. due to reheating by r-modes.  For the kaon-kaon process,  assuming a toy interaction Lagrangian with a single kaon coupling constant (instead of the usual three), it is predicted that $\eta$ depends microscopically on the Goldstone kaon phase speed and on whether the scattering occurs via a contact process or the exchange of a virtual particle.  From Fig. 2 in \citet{alf09}, we see that the kaon-kaon viscosity is between $\sim10^{12}$ and $\sim10^{15}$ times lower than the phonon-phonon viscosity given by (\ref{goeq36}), since the kaon chemical potential $(\approx4\,{\rm MeV})$ is much less than $\mu_q$.  Importantly, however, for a typical core temperature $T\sim10^8\,{\rm K}$, the phonon-phonon mean free path exceeds the stellar radius, so the kaons are the leading residual source of viscosity, with $10^{-25}\la E \la 10^{-16}$.  The upper end of this range is consistent with the Crab and Vela data in Figure \ref{fig9}.


\citet{man09b} showed that, for elastic phonon-vortex scattering (inelastic is suppressed) in the CFL phase, the mutual friction depends of the speed of sound, $c_s\approx3^{-1/2} c$, and the phonon density, with
\begin{equation}\label{goeq37}
 B=\frac{2 \pi^2 c^6}{405 c_s^6}\left(1-\frac{c_s^2}{c^2}\right)\left(\frac{k_B T}{\mu_q}\right)^5\,.
\end{equation}
For densities and temperatures in the core (see above), (\ref{goeq37}) evaluates to $5\times10^{-21}\la B\la 5 \times 10^{-16}$.  This is several orders of magnitude smaller than the typical $B$ value extracted from the model in \S\ref{gosec3d} and \S\ref{gosec4d}.

In the 2SC phase, the $u$ and $d$ quarks pair up to form a condensate and two quark colors acquire a gap $(\Delta \sim 50\,{\rm MeV})$.
However, one quark color remains ungapped, and this color inevitably dominates all dissipative processes in the system; the contribution to transport by the gapped colors is exponentially suppressed $\propto{\rm exp}(-\Delta/3 k_B T)$, as in the CFL phase.
(As no global symmetries are broken, the 2SC phase is also not a superfluid.)
Detailed enumeration of the interaction channels available to the ungapped quark \citep{mad00, sad08} indicates that the shear viscosity of the 2SC phase is $(5/9)^{1/3}$ times the shear viscosity for standard, ungapped quark matter (with no diquark pairing), i.e., $\eta$ is $(5/9)^{1/3}$ times equation (\ref{goeq35}).
For densities and temperatures in the core (see above), this implies $4\times10^{-15} \la E\la 5\times10^{-13} $.
Our conclusions with respect to the glitch data are therefore the same, namely that the CFL phase (like standard, ungapped quark matter) is broadly consistent with the observations, especially at the lower end of the range.
Ab initio theoretical calculations of $B$ in the 2SC phase have not yet been published, to the best of our knowledge.

\subsection{Depth averages}

\begin{figure}
 \includegraphics[width=74mm]{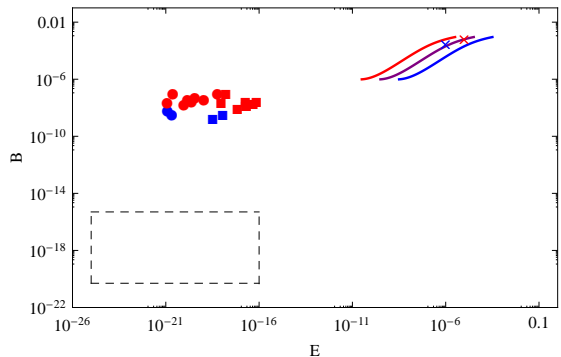}
 \caption{The three curves are obtained by varying the density $\rho$ for three different values of temperature, $T=10^{7.5}\rm{K}$  (top, red), $T=10^{7.0}\rm{K}$  (centre, green) and $T=10^{6.5}\rm{K}$  (bottom, blue). Squares represent fits with $K=50$ and while circles are for $K=1$.
The Vela fits are shown in red, and the Crab is shown in blue. }
\label{fig12}
\end{figure}

An interesting property of the Crab and Vela fits is that they naturally yield values of the product $B E^{-1/2}$ and $\rho_n \left(1+K\right)$ which agree nicely with nuclear theory, e.g. in Figures \ref{fig9} and \ref{fig12}, yet $B$ and $E$ disagree individually with nuclear predictions.  One possibility, of course, is that the agreement in $B E^{-1/2}$ and $\rho_n\left(1+K\right)$ is accidental; that is, the results are telling us that a self-consistent but purely hydrodynamic model is incapable of explaining the relaxation data, and new, non-hydrodynamic physics must be introduced.  If true, this would fulfill one of the papers chief aims, as discussed in \S\ref{gosec1}.  However, given that the discrepancies in $B$ and $E$ are not specific to this particular relaxation model but are the generic consequences of a hydrodynamic treatment, there is another, equally likely explanation: stratification and compressibility restrict Ekman pumping to a thin surface layer [not treated in this paper; cf. \citet{abn96} and \citet{van09}], so that the effective, body-averaged values of $B$ and $E$ used in the model are very different to what one would expect for a uniform-density star.  Figures \ref{fig9} and \ref{fig12} amply demonstrate this: the products $B E^{-1/2}$ and $\rho_n\left(K+1\right)$ depend weakly on density, varying $\sim30$-fold over the range $10^{12}\leq\rho/\rm{g\,cm^{-3}}\leq 10^{17}$ (Figure \ref{fig9}), but individually they vary $\sim10^5$-fold over the same range of $\rho$.  Furthermore, to bring the $B$ and $E$  fits into closer agreement with nuclear theory, one would need to shift the neutron-rich-matter curves in Figure \ref{fig12} to the left and down, just as one expects if the Ekman layer is restricted.

Figure \ref{fig12} displays the fitted values of $B$ and $E$ from \S\ref{gosec3} and \S\ref{gosec4} for all Crab (blue points) and Vela (red points) glitches, assuming $K=1$ (circles) and $K=50$ (squares).  Overall, both $B$ and $E$ are smaller than the theoretical predictions in (\ref{goeq31})--(\ref{goeq33}) for neutron-rich matter. The coupling time $\tau_\nu\sim 1\,\rm{s}$ determined by \citet{alp84} for typical values of density, proton density fraction and effective proton mass translates to $B\sim3\times10^{-4}$ and $B\sim7\times10^{-4}$ using Mendell's relation [see Eq (21) in \citet{men91b}] for the Crab and Vela respectively.  Similarly, the respective Ekman times estimated by \citet{eas79} are translate to $E\sim 10^{-6}$ and $10^{-5}$, marked as blue and red crosses in Figure \ref{fig12}. Compositional variations such as the existence of strange quark matter in the interior also affect the depth-averaged effective parameters, as discussed in \S\ref{gosec2a}. The ranges of $B$ and $E$ for the CFL phase from \S\ref{gosec5b} are sketched on figure \ref{fig12} as a dashed rectangle. They are uncertain, of course, but depth-averaging over strange quark matter dramatically reduces the effective $B$ and $E$. Figure \ref{fig12} illustrates that the self-consistent hydrodynamic model potentially possesses considerable discriminatory power between nuclear models which are widely separated on the $B$--$E$ plane, once it is refined further.

\subsection{Crust fraction}\label{gosec5c}

The theoretical curves in Figures \ref{fig9} and \ref{fig10} are drawn for the canonical value $K=50$ of the crust-fluid fraction.  How does this compare with terrestrial experiments that probe the equation of state of neutron-rich nuclear matter?  \citet{lat07} showed that $K$ depends sensitively on the symmetry energy at subsaturation densities through the transition pressure $p_+$ and density $\rho_+$ at the solidification point.  Quantitatively, one has
\begin{equation} \label{goeq38}
 \frac{1}{K}=\frac{28 \pi p_+ R_*^3}{3 M_* c^2}\frac{\left(1-1.67 \xi-0.6 \xi^2\right)}{\xi}\left[1+\frac{2 p_+\left(1+5\xi-14\xi^2\right)}{\rho_+m_bc^2\xi^2}\right]^{-1}
\end{equation}
with $\xi=G M_*/R_* c^2$, where $m_b$, $M_*$, and $R_*$ denote the mean baryon mass, stellar mass, and stellar radius respectively.  Recently, \citet{xu09} analysed isospin diffusion data from heavy-ion collisions and placed bounds on the density dependence of the symmetry energy.  They concluded that $p_+$ and $\rho_+$ lie in the ranges $0.01\,{\rm MeVfm^{-3}}\leq p_+\leq 0.26\, {\rm MeVfm^{-3}}$ and $0.040\,{\rm fm^{-3}}\leq \rho_+/m_n \leq 0.065\, {\rm MeVfm^{-3}}$, consistent with the isotopic dependence of giant monopole resonances in \rm{Sn} \citep{li07} and the neutron-skin thickness of $^{208}$\rm{Pb} \citep{li05}.  The above ranges imply $86 \leq K \leq 1779 $.  Again, this prediction is consistent with the results in Figure \ref{fig9}, allowing for moderate shifts of the theoretical curves to the right.  It is also  consistent with the bound $K<70$ inferred from Vela glitches \citep{lin99} invoking angular momentum conservation.


\section{Conclusions}  \label{gosec6}
A two-fluid hydrodynamic model which predicts the spin evolution of the crust during the recovery stage of a pulsar glitch is presented.  The model has six free parameters ($\rho_n$, $K$, $B$, $E$, $\Omega_{n0}$ and $\Omega_0$) and solves for the back reaction of the viscous torque on the crust in a self-consistent manner.
This analysis represents the next step in the class of models introduced by \citet{sid09}, by replacing body-averaged internal components with an exact global Ekman flow pattern.

The paper presents a straightforward recipe for extracting two important transport coefficients in bulk nuclear matter, namely the shear viscosity and mutual friction parameter, from radio timing observations of glitch recovery.  Given a post-glitch timing solution of the multi-exponential form (\ref{goeq17}), or equivalently (\ref{goeq19}), one fits the data to the theoretical expression (\ref{goeq22}) and reads off four of the model parameters (e.g. $B$, $E$, $\Omega_{n0}$ and $\Omega_0$) in terms of the other two (e.g., $K$, $\rho_n$) from equations (\ref{goeq20}), (\ref{goeq21}) and (\ref{goeq23})--(\ref{goeq25}).  The fits (e.g. in Figures \ref{figvela85a}, \ref{figvela88} and \ref{figcrab75}) are excellent and include a natural explanation of the ``overshoot'' phenomenon seen in the Crab.  For glitches with three resolvable time-scales, the model parameters are constrained uniquely, and satisfy $(7 B E^{-1/2})/(20 \rho_n K)<1$.   This implies that the viscous time-scale, appearing in (\ref{goeq25}), is shorter than the mutual friction time-scale, appearing in (\ref{goeq24}).  When the fits derived for all glitches with $N\geq2$ in Vela (Figure \ref{figvelascat}) and the Crab (Figure \ref{figcrabscat}) are plotted together on the $B E^{-1/2}$-$\rho_n (K+1)$ plane, the regions covered by the two objects overlap.  Moreover, the region of overlap coincides with curves of $B E^{-1/2}$ versus $\rho_n (K+1)$ predicted from nuclear theory, both for a standard $^1S_0$ neutron superfluid at the temperatures and densities expected in the outer core and also for certain phases of strange quark matter (e.g. stable ungapped phase, 2SC phase, or CFL phase with kaon viscosity) under inner core conditions.  The crust-fluid ratio inferred from the same fits is consistent with recent measurements of isospin diffusion in heavy-ion collisions \citep{xu09}.

The self-consistent spin-down model in this paper helps solve two of the puzzles raised in \S\ref{gosec1}, namely that glitch recovery cannot be parameterized by a simple exponential decay and that it is not always monotonic.  It does not solve the third puzzle -- that the recovery time-scale varies from glitch to glitch in the same object -- but it does shed new light upon it by sharpening the debate.  Of the six parameters, the transport coefficients $B$ and $E$ do not change significantly during the interval between glitches.  The inertia-related factors $\rho_n$ and $K$ may conceivably change with each event, if different zones within the star (e.g. hosting different average pinning strengths) participate in different glitches \citep{alp93,alp96, sed02}.  However, the global nature of the Ekman process argues against this; even if successive glitches are triggered in distinct pinning zones, the resulting global flow embraces the entire neutron condensate and charged fluid (and exerts a torque on the whole crust), so that the effective values of $\rho_n$ and $K$ are always the same.  This leaves one possibility: $\Omega_{n0}$ and $\Omega_0$ vary from glitch to glitch.  Such an outcome is natural if different pinning zones trigger different glitches; the shear preceding a glitch is not always the same (e.g.  we do not observe a reservoir effect in the data).

The model presented in this paper possesses several serious weaknesses.  First, it fails to explain the fourth time-scale $t_4\sim{\rm min}$ measured by \citet{dod02,dod07} in two Vela glitches. This phenomenon is probably a result of physics not included in this analysis, e.g., the relaxation of magnetic flux tubes \citep{van08, alp09}, or a solid core in a superconducting phase.  Second, the two-fluid theory neglects entrainment \citep{and06b,sid09}, to keep the difficult spin-down problem tractable analytically, even though entrainment is known to be important.  Our conclusions regarding $\eta$ and $B$ are expected to change by factors of a few due to this effect.
Third, the analysis neglects bulk viscosity, because it enters at order $O(E)$ in the equations of motion and therefore does not modify the Ekman process.  In practice, however, the bulk viscosity exceeds the shear viscosity by several orders of magnitude in several phases of neutron-rich and exotic quark matter as when electroweak processes like $d+s\leftrightarrow u+s$ work to restore chemical equilibrium \citep{mad00,man05,jai08,sad08}.
Lastly, although the values of $B E^{-1/2}$ extracted when fitting the model to data are consistent with what nuclear theories predict, the individual values of $B$ and $E$ are not.
The most natural explanation is that $B$ and $E$  depend sensitively on $\rho$ whereas the combination $B E^{-1/2}$ does not, and that stratification and compressibility effects limit the Ekman flow to a narrow surface layer in a way that is not modeled properly in this paper.  This important issue deserves further study.

More radio timing data on glitch recoveries is urgently required to fill out Figure \ref{fig9} and test the theory more comprehensively.  Multibeam monitoring experiments with phased radio arrays are under consideration and offer exciting prospects in this regard.

\section*{Acknowledgments}

CAVE acknowledges the financial support of an Australian Postgraduate Award.

\bibliographystyle{mn2e}
\bibliography{glitchobs}

\appendix

\section[]{General solution to the two-fluid Ekman problem}

The general solution to (\ref{goeq1})--(\ref{goeq4}) in an arbitrary container for any choice of $B,B',E,\rho_n$ is given by \citet{van10}
\begin{equation} \label{goeqa01}
v_{\phi}(r,\tau)=\frac{r \omega_+(r) \omega_-(r)}{\beta \left[\omega_+(r)-\omega_-(r)\right]} \int_0^\tau\,\rmn{d}\tau' f(\tau')
\end{equation}
\[
\phantom{v_{\phi}(r,\tau)=} \times\left\{  \left[\omega_+(r)+\beta\right] e^{\omega_+(r)\left(\tau-\tau'\right)} \right.
\]\[
\phantom{v_{\phi}(r,\tau)=} \left. -\left[\omega_-(r)+\beta \right]e^{\omega_-(r)\left(\tau-\tau'\right)}\right\} \nonumber
\]\[
\phantom{v_{\phi}(r,\tau)=} -\frac{r \Omega_{n0} \omega_+(r) \omega_-(r) }{\beta\left[\omega_+(r)-\omega_-(r)\right]} \left[e^{\omega_+(r) \tau}- e^{\omega_-(r) \tau}\right] \nonumber
\]\[
\phantom{v_{\phi}(r,\tau)=} -\frac{r \Omega_0}{\omega_+(r)-\omega_-(r)}\left[\omega_-(r)e^{\omega_+(r) \tau}-\omega_+(r) e^{\omega_-(r) \tau}\right] \, .
\]
where we have defined
\begin{equation}\label{goeqa02}
 \beta=\frac{2 B E^{-1/2}}{2-B'} \, ,
\end{equation}
which characterises the relative strength of the superfluid mutual friction force to the viscous forces, and the exponential time-scales
\begin{equation} \label{goeqa03}
  \omega_{\pm}(r)=-\frac{1}{2}\left[\beta+\frac{I(r)}{h(r)}\right]\pm \sqrt{\frac{1}{4}\left[\beta+\frac{I(r)}{h(r)}\right]^2-\frac{\beta \rho_n J(r)}{h(r)}} \, .
\end{equation}
which are a combination of the classical Ekman time-scales and the mutual friction time-scale.  In (\ref{goeqa03}), we have
\begin{equation}  \label{goeqa04}
 I(r)=\frac{1}{\lambda_{-}(r)\left[\lambda_{-}^2(r)+\lambda_{+}^2(r)\right]  }
\end{equation}
\[ \phantom{I(r)=} \times \{\left[\frac{1-H^4(r)}{1+H^4(r)}\right]^2\left[\lambda_{+}^2(r)-\lambda_{-}^2(r)\right]^2+4\lambda_{+}^2(r)\lambda_{-}^2(r)\}^{1/2}\, ,
\]
\begin{equation}\label{goeqa05}
 J(r)=\frac{H^2(r)}{2\lambda_{-}(r) \left[1+H^4(r)\right]}
\end{equation}
\[ \phantom{J(r)=} \times \left\{ \left[\lambda_{-}^2(r)+\lambda_{+}^2(r)\right]\left[1-H^4(r)\right]+4 \lambda_{-}^2(r) H^4(r)\right\} \, ,
\]
and
\begin{equation}\label{goeqa06}
 \lambda_{\pm}(r)=\frac{1}{H(r)}\left\{ \left[\frac{\left(B'-2\right)^2+B^2}{\left(\rho_n B'-2\right)^2+\left(\rho_n B\right)^2}\right]^{1/2} \right.
\end{equation}
\[
 \phantom{\lambda_{\pm}(r)=}\left. \mp \frac{\rho_s B\left(1+H^4(r)\right)}{H^2(r)\left[\left(\rho_n B'-2\right)^2+\left(\rho_n B\right)^2\right]}\right\}^{1/2} \, .
\]
The factors $H(r)$ and $h(r)$ are related to the shape of the boundary.  In a general container which is symmetric about the $z=0$ plane, at the upper boundary $z=h(r)$ while at the lower boundary $z=-h(r)$.  Then $H(r)$ is defined as
\begin{equation}\label{goeqa07}
 H(r)=\left\{1+\left[h'(r)\right]^2\right\}^{1/4} \, ,
\end{equation}
For a sphere, they are given by \citep{van10}
\begin{equation} \label{goeqa08}
 h(r)=H(r)^{-2}=\left(1-r^2\right)^{1/2} \, ,
\end{equation}

The viscous torque on the crust is computed from the right hand side of
\begin{equation} \label{goeqa09}
 \frac{\partial f(\tau)}{\partial \tau}= -\frac{15 K}{2}\int_{0}^{1}\rmn{d}r\,r^2 h(r)\frac{\partial v_{\phi}(r)}{\partial \tau}  \, .
\end{equation}
Equations (\ref{goeqa01}) and (\ref{goeqa09}), combine to give an integral equation for the spin evolution of the crust, viz.
\begin{equation}\label{goeqa10}
 f(\tau)=-\rho_n K \int_0^\tau \rmn{d}\tau' \, \left[\dot{g}^A(\tau-\tau')+\dot{g}^B(\tau-\tau')\right]f(\tau')
\end{equation}
\[\phantom{ f(\tau)=} + \rho_n K  \left[  g^A(\tau)\Omega_{n0}+ g^B(\tau) \Omega_{0} \right] +1 \, ,
\]
where
\begin{equation} \label{goeqa11}
 g^A(\tau)=-\frac{15}{2}\int_{0}^{1} \rmn{d}r \, \frac{r^3 J(r)\left[  e^{\omega_+(r)\tau}-e^{\omega_-(r)\tau} \right]}{\omega_-(r)-\omega_+(r)}
\end{equation}
\begin{equation} \label{goeqa12}
 g^B(\tau)=\beta \int_0^\tau \rmn{d}\tau' \, g^A(\tau') \, .
\end{equation}

In the limit $B,B'\ll1$ (but not necessarily $B E^{-1/2}\ll1$), we find that (\ref{goeqa02})--(\ref{goeqa06}) simplify to
\begin{equation} \label{goeqa13}
 \beta=B E^{-1/2}
\end{equation}
\begin{equation} \label{goeqa14}
 I(r)=J(r)=\lambda_{\pm}^{-1}(r)= H(r)
\end{equation}
which give the results (\ref{goeq11})--(\ref{goeq16}).

\section[]{General solution of the one-fluid Ekman problem} \label{gosecApB}

When there is no superfluid component, we have $B=B'=0$, $\rho_n=1$, $\Omega_{n0}=\Omega_0$.  Therefore $I(r)=J(r)=\lambda_{\pm}^{-1}(r)= H(r)$,
$\beta=0$, $\omega_+=0$, and $\omega_-=-H(r)/h(r)$.  Equation (\ref{goeqa01}) then reduces to
\begin{equation} \label{goeqb01}
v_{\phi}(r,\tau)=\frac{r H(r)}{h(r)} \int_0^\tau \, \rmn{d}\tau' f(\tau') e^{-\frac{ H(r)}{h(r)}\left(\tau-\tau'\right)}+r \Omega_0 e^{-\frac{ H(r)\tau}{h(r)}}
\end{equation}
which is the classical solution obtained by \citet{gre63}.
In a sphere, the boundary is defined by (\ref{goeqa08}).
Equations (\ref{goeqa10})--(\ref{goeqa12}) then become
\begin{equation}\label{goeqb02}
 f(\tau)=- K \int_0^\tau \rmn{d}\tau' \, \dot{g}^A(\tau-\tau')f(\tau')
\end{equation}
\[\phantom{ f(\tau)=} +  K  g^A(\tau)\Omega_{0} +1 \, ,
\]
\begin{equation} \label{goeqb03}
 g^A(\tau)=-\frac{15}{2}\int_{0}^{1} \rmn{d}r \, r^3 h(r) \left[e^{-\frac{H(r)}{h(r)}\tau}-1 \right]
\end{equation}
\begin{equation} \label{goeqb04}
 g^B(\tau)=0 .
\end{equation}
The only variables in this limit are $K$, $\Omega_0$,  and $E$.
Equations (\ref{goeq20}) and (\ref{goeq21}) simplify to
\begin{equation} \label{goeqb05}
\frac{\Delta\nu_p}{\Delta\nu}=\frac{1+K \Omega_0}{1+K} \, ,
\end{equation}
\begin{eqnarray} \label{goeqb06}
\frac{\sum_{n=1}^N \Delta \nu_n/t_n}{2\pi \nu_{g0} \Delta\nu} =\frac{20}{7} K E^{1/2} \left(1-\Omega_{0}\right)\, ,
\end{eqnarray}
providing two conditions in terms of the observed quantities on the left-hand sides.
We are left with one model variable (say, $K$) which is then determined by fitting the shape of $f(\tau)$ to $f_{obs}(\tau)$ by eye.

\section[]{Parameters extracted from by-eye fits}
Table \ref{tab3} quotes the model parameters $\rho_n$, $K$, $B$, $E$, $\Omega_{0}$ and $\Omega_{n0}$ for all glitches fitted by-eye.

\begin{table*} \label{tab3}
 \centering
 \begin{minipage}{1500mm}
  \caption{By-eye model fits to glitch data}
  \begin{tabular}{@{}ccccccccc@{}}
  \hline
   Object & Glitch & Ref.&  $\rho_n$ & $K$ & $B \times 10^{-10} $ & $E \times 10^{-20} $ & $\Omega_{0}$ & $\Omega_{n0}$  \\
 \hline
Crab & 1975  & 1 & 0.1 & 80 & 8.5 & 1.30 & 0.77 & 0.127 \\
 & &   & 0.1 & 500 & 7.0 & 0.049 & 0.77 & 0.28 \\
 & &   & 0.1 & 750 & 6.5 & 0.024 & 0.77 & 0.31 \\
 & &   & 0.1 & 1000 & 6.36 & 0.014 & 0.77 & 0.32 \\
 & &   & 0.1 & 1500 & 6.36 & 0.0062 & 0.77 & 0.32 \\
 & &   & 0.1 & 2500 & 6.36 & 0.0023 & 0.77 & 0.33 \\
&  &   & 0.01 & 2500 & 7.3 & 0.18 & 0.77 & 0.25 \\
 & &   & 0.1 & 80 & 32 & 0.083 & 0.77 & -2.46 \\
 & &   & 0.1 & 500 & 33.5 & 0.0020 & 0.77 & -2.55 \\
&  &   & 0.1 & 750 & 34 & 0.00087 & 0.77 & -2.59 \\
&  &   & 0.1 & 1000 & 34 & 0.00049 & 0.77 & -2.59 \\
&  &   & 0.1 & 1500 & 34 & 0.00022 & 0.77 & -2.59 \\
&  &   & 0.1 & 2500 & 34 & 0.00008 & 0.77 & -2.59 \\
&  &   & 0.01 & 2500 & 33 & 0.0084 & 0.77 & -2.47 \\
\hline
Vela & 1985 & 2  & 0.1 & 1 & 228 & 92.8 & 0.68 & 0.38 \\
 & &   & 0.1 & 50 & 252 & 0.116 & 0.84 & 0.65 \\
 & &   & 0.001 & 500 & 253 & 14.7 & 0.84 & 0.69 \\
 & &   & 0.001 & 2000 & 253 & 0.78 & 0.84 & 0.67 \\
 & &   & 0.001 & 5000 & 253 & 0.13 & 0.84 & 0.67 \\
 & &   & 0.01 & 2000 & 253 & 0.0078 & 0.84 & 0.67 \\
 & &   & 0.01 & 5000 & 253 & 0.0013 & 0.84 & 0.67 \\
\hline
Vela & 1988 & 3  & 0.01 & 53 & 50 & 3.05 $\times 10^5$ & 0.97 & 0.97 \\
\hline

\multicolumn{7}{l}{[1] \citep{lyn93}, [2] \citep{mcc87}, [3] \citep{fla90}}
\end{tabular}
\end{minipage}
\end{table*}

\bsp

\label{lastpage}

\end{document}